\documentclass[american,aps,pra,reprint,longbibliography,superscriptaddress]{revtex4-2}

\usepackage{lmodern}
\usepackage[T1]{fontenc}
\usepackage[utf8]{inputenc}
\usepackage{color}
\usepackage{babel}
\usepackage{units}
\usepackage{amsmath}
\usepackage{amssymb}
\usepackage{graphicx}
\usepackage[pdfusetitle,
 bookmarks=true,bookmarksnumbered=false,bookmarksopen=false,
 breaklinks=false,pdfborder={0 0 0},pdfborderstyle={},backref=false,colorlinks=true]
 {hyperref}
\hypersetup{
 citecolor=blue,linkcolor=teal,urlcolor=blue}

\makeatletter

\usepackage{xcolor}

\makeatother

\begin{document}
\global\long\def\i{\mathrm{i}}%
\global\long\def\d{\mathrm{d}}%
\global\long\def\e{\mathrm{e}}%
\global\long\def\Tr{\mathrm{Tr}}%
\global\long\def\bra#1{\langle#1|}%
\global\long\def\ket#1{|#1\rangle}%
\global\long\def\braket#1#2{\left\langle #1|#2\right\rangle }%
\global\long\def\ketbra#1#2{\left|#1\right\rangle \!\left\langle #2\right|}%
\global\long\def\dbra#1{\langle\!\langle#1|}%
\global\long\def\dket#1{|#1\rangle\!\rangle}%
\global\long\def\dbraket#1#2{\left\langle \!\left\langle #1|#2\right\rangle \!\right\rangle }%
\global\long\def\dketbra#1#2{\left.\!\left|#1\right\rangle \!\right\rangle \!\left\langle \!\left\langle #2\right|\!\right.}%

\title{Light-induced localized vortices in multicomponent Bose--Einstein condensates}

\author{Y.~Braver}
\affiliation{Institute of Theoretical Physics and Astronomy, Faculty of Physics,
Vilnius University, Saulėtekio 3, LT-10257, Vilnius, Lithuania}
\author{D.~Burba}
\affiliation{Institute of Theoretical Physics and Astronomy, Faculty of Physics,
Vilnius University, Saulėtekio 3, LT-10257, Vilnius, Lithuania}
\author{S.~S.~Nair}
\affiliation{Quantum Systems Unit, Okinawa Institute of Science and Technology
Graduate University, 904-0495 Okinawa, Japan}
\author{G.~Žlabys}
\affiliation{Quantum Systems Unit, Okinawa Institute of Science and Technology
Graduate University, 904-0495 Okinawa, Japan}
\author{E.~Anisimovas}
\affiliation{Institute of Theoretical Physics and Astronomy, Faculty of Physics,
Vilnius University, Saulėtekio 3, LT-10257, Vilnius, Lithuania}
\author{Th.~Busch}
\affiliation{Quantum Systems Unit, Okinawa Institute of Science and Technology
Graduate University, 904-0495 Okinawa, Japan}
\author{G.~Juzeliūnas}
\affiliation{Institute of Theoretical Physics and Astronomy, Faculty of Physics,
Vilnius University, Saulėtekio 3, LT-10257, Vilnius, Lithuania}
\begin{abstract}
We study continuous interaction of a trapped two-component Bose--Einstein condensate with light fields in a $\Lambda$-type configuration. Using light beams with orbital angular momentum, we theoretically show how to create a stable, pinned vortex configuration, where the
rotating component is confined to the region surrounded by the second,
non-rotating component. The atoms constituting this vortex can be
localized in volumes much smaller than the volume occupied by the
second component. We also show that the vortex position can be changed dynamically by moving the laser beams, provided the beams' movement speed remains below the speed of sound. This allows us to use the localized vortex to stir the second component, and to determine the superfluid flow's critical velocity.

\end{abstract}
\maketitle

\section{Introduction}

Bose--Einstein condensates (BECs) provide a unique platform for studying
quantum fluids, which can support macroscopic quantum phenomena such as
superfluidity and quantized vortices. Vortices in BECs are
characterized by phase singularities in the condensate wave function
and are fundamental excitations that reveal insights into angular
momentum quantization and topological defects in quantum systems \citep{Fetter2001}.
The study of vortices has broad implications, ranging from quantum
turbulence \citep{Tsubota2013,Reeves2022} to connections with superconductivity
\citep{Harada1992} and the structure of neutron stars \citep{Magierski2024}.

A key challenge in the study of BEC vortices is their controlled creation
and manipulation. Typically, the vortices are produced by transferring angular momentum to the atoms using optical
means: through phase imprinting \citep{Dobrek1999}, stirring \citep{Madison2000,AboShaeer2001},
or using beams carrying orbital angular momentum, such as Laguerre--Gaussian (LG)
beams \citep{Marzlin1997,Dum1998,Nandi2004,Kapale2005,Juzeliunas2005Meissner,Andersen2006,Wright2008,Wright2009,Mukherjee2021}.
More recently, interest has also increased in coupling light to condensates using the so-called $\Lambda$-type configurations, which create an internal dark state. The properties of this state can be used to realize atom control at subwavelength
resolution \citep{Juzeliunas2007,Gorshkov2008}, create optical lattices featuring
barriers of subwavelength width \citep{Lacki2016,Gvozdiovas2023},
and make narrow structures in the BEC \citep{Hamedi2022}. The $\Lambda$-systems
have also been used to create vortices in BEC via time-dependent transfer of population between the atomic internal states by applying Raman-type schemes
\citep{Nandi2004,Dutton2004,Kapale2005,Thanvanthri2008}.

In this work, we consider a trapped two-component BEC continuously interacting with two light fields in a $\Lambda$-type configuration. One of the two laser fields used in the setup is an LG beam, which can transfer angular momentum to the BEC atoms.
We show in the following that the driven system possesses stationary states such that either one or both components of the BEC are in a vortex state. These stationary states belong to the manifold of dark states and are therefore immune to decay via spontaneous emission. Compared to the familiar coreless-vortex profiles (whereby the core of the rotating vortex is filled by the non-rotating component), the stationary states can also have ``inverted'' profiles, with the rotating component being surrounded by the non-rotating one. In this case, the rotating atoms of one component are tightly localized in a region of the order of the healing length of the other component. The degree of localization can be controlled by tuning the relative strength of the two laser beams. Remarkably, when this relative strength exceeds a certain threshold, the localized-vortex state becomes the lowest-energy state of the dark-state manifold. To further demonstrate the stability of these states, we show that the resulting vortices can be robustly moved around the trap by moving the laser beams. As long as the beam movement speed is below the speed of sound, the vortex follows the beams without notably disturbing the non-rotating component.
To quantify this effect, we move the vortex in circular paths, thereby stirring the other component with a localized impurity. By examining the resulting heating, we determine the critical velocity of the superfluid flow below which the vortex moves without dissipation.

\section{Theory}

\subsection{Single-particle Hamiltonian of the system\label{subsec:Hamiltonian}}

\begin{figure}
\begin{centering}
\includegraphics{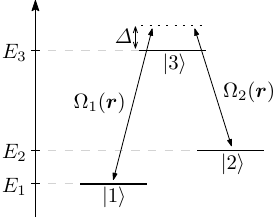}
\par\end{centering}
\caption{\protect\label{fig:levels}Schematic representation of the relevant
energy levels of the $\Lambda$ scheme of atom-light coupling.}
\end{figure}

We consider a general three-level atomic system arranged in a $\Lambda$-type
configuration of the atom-light coupling shown in Fig.~\ref{fig:levels}
\citep{Dum1996,Juzeliunas2004,Juzeliunas2005,Juzeliunas2005EIT,Juzeliunas2007,Gorshkov2008,Lacki2016,Jendrzejewski2016}.
Two co-propagating light beams provide the position-dependent couplings
between the atomic internal states. The beams are characterized by
the Rabi frequencies:
\begin{equation}
\begin{split}\Omega_{1}(\rho,\varphi) & =\Omega_{0}\left(\frac{\rho}{a}\right)^{\nu}\e^{-\rho^{2}/w_{0}^{2}}\e^{\i k_{z}z+\i\nu\varphi},\\
\Omega_{2}(\rho,\varphi) & =\Omega_{0}\e^{-\rho^{2}/w_{0}^{2}}\e^{\i k_{z}z},
\end{split}
\label{eq:beams}
\end{equation} 
where $(\rho,\varphi)$ are the polar coordinates, $w_{0}$ is the
beam waist (the same for both beams), $\Omega_{0}$ is the amplitude
of the Rabi frequencies, and $k_{z}$ is the wave number for the paraxial
propagation of the light beams in the $z$ direction. The parameter
$a$, having dimension of length, controls the subwavelength nature
of the setup. The beam characterized by the Rabi frequency $\Omega_{1}$
couples the atomic internal states $\ket 1$ and $\ket 3$ and represents
a Laguerre--Gaussian mode ${\rm LG}_{0}^{\nu}$ with a vorticity (winding
number) $\nu>0$. On the other hand, the atomic states $\ket 2$
and $\ket 3$ are coupled by a Gaussian beam characterized by the
Rabi frequency $\Omega_{2}$. 
Both beams co-propagate along the $z$-axis \citep{Juzeliunas2004,Juzeliunas2005EIT,Juzeliunas2005},
oscillating at frequencies $\omega_{1}$ and $\omega_{2}$, respectively.

On a single-particle level, the dynamics of an atom is governed by the Hamiltonian
\begin{equation}
\hat{H}(\boldsymbol{r})=-\frac{\hbar^{2}}{2m}\Delta+V(\boldsymbol{r})+\hat{H}_{0}(\boldsymbol{r}).\label{eq:H}
\end{equation}
The kinetic energy term and the trapping potential $V(\boldsymbol{r})$
(assumed to be the same for the atomic internal states  $\ket 1$, $\ket 2$ and $\ket 3$)
act as identity operators in the space of atom's internal states.
Transitions between the latter states are described by the internal-space
Hamiltonian
\begin{equation}
\begin{split}\hat{H}_{0}(\boldsymbol{r})  =\,&\epsilon\ketbra 22+(-\varDelta-\i\tfrac{\Gamma}{2})\ketbra 33\\
 & +\hbar[\Omega_{1}(\boldsymbol{r})\ketbra 31+\Omega_{2}(\boldsymbol{r})\ketbra 32+{\rm H.c.}].
\end{split}
\label{eq:H0}
\end{equation}
The temporal dependence of this Hamiltonian has been eliminated by transitioning to the rotating frame and adopting the rotating wave approximation \cite{ScullyZubairyBook}. Two parameters emerge as a result, including the single-photon
detuning $-\varDelta=E_{3}-E_{1}-\hbar\omega_{1}$ and the two-photon
detuning $\epsilon=E_{2}-E_{1}+\hbar\omega_{2}-\hbar\omega_{1}$;
here $E_{i}$ denotes the energy of the $i$th energy level (see Fig.~\ref{fig:levels}).
Furthermore, we have introduced the spontaneous decay rate $\Gamma$
for the excited state $\ket 3$ represented through the imaginary part of the excited-state energy. In the following, we consider the two-photon resonance, for which $\epsilon=0$.

It is instructive to consider the states $\ket{\alpha(\boldsymbol{r})}$
representing the eigenstates of the internal-space Hamiltonian (\ref{eq:H0}),
such that $\hat{H}_{0}(\boldsymbol{r})\ket{\alpha(\boldsymbol{r})}=\varepsilon_{\alpha}(\boldsymbol{r})\ket{\alpha(\boldsymbol{r})}$,
with $\boldsymbol{r}$ treated parametrically. Of special interest
is the ``dark'' eigenstate of $\hat{H}_{0}$, given by:
\begin{equation}
\ket{{\rm D}(\boldsymbol{r})}=\frac{1}{\sqrt{1+|\zeta|^{2}}}(\ket 1-\zeta\ket 2)\,,
\end{equation}
where 
\begin{equation}
\zeta(\boldsymbol{r})=\frac{\Omega_{1}(\boldsymbol{r})}{\Omega_{2}(\boldsymbol{r})}.\label{eq:zeta}
\end{equation}
The dark state is characterized by zero eigenenergy. Having no contribution
from the lossy excited state $\ket 3$, the dark state has an infinite
lifetime as far as evolution under $\hat{H}_{0}$ is concerned. For
$\varDelta=\Gamma=0$, the remaining two eigenstates of the internal-space
Hamiltonian are given by 
\begin{equation}
\ket{\pm}=\frac{1}{\sqrt{2}}\left(\ket{{\rm B}}\pm\ket 3\right)\,,\quad\mathrm{where}\quad\ket{{\rm B}}=\frac{\zeta^{*}\ket 1+\ket 2}{\sqrt{1+|\zeta|^{2}}}\,,\label{eq:pm-states}
\end{equation}
with 
\begin{equation}
\hat{H}_{0}\ket{\pm}=\pm\hbar\Omega\ket{\pm}\quad\mathrm{and}\quad\Omega=\sqrt{\left|\Omega_{1}^{2}\right|+\left|\Omega_{2}^{2}\right|}\,.\label{eq:pm-eigenvalue_eq}
\end{equation}
In Eq.~(\ref{eq:pm-states}), $\ket{{\rm B}}$ is the so-called bright
state, representing a superposition of atomic ground states orthogonal
to the dark state $\ket{{\rm D}}$. Later we will consider the adiabatic
motion of atoms in the dark-state manifold, which is relevant if the
total Rabi frequency $\Omega$ is much larger than the characteristic
kinetic energy of the atomic center of mass motion.

\subsection{Dynamics of the BEC}

Now let us consider the equations for an atomic BEC interacting with the laser fields. We start with the Schrödinger equation for the full state vector $\ket{\Phi(\boldsymbol{r})}$ of a single atom
\begin{equation}
\i\hbar\frac{\partial}{\partial t}\ket{\Phi(\boldsymbol{r})}=\hat{H}(\boldsymbol{r})\ket{\Phi(\boldsymbol{r})}
\label{Schroedinger-eq}
\end{equation}
and use the expansion 
\begin{equation}
\ket{\Phi(\boldsymbol{r})}=\sum_{i=1}^{3}\Phi_{i}(\boldsymbol{r})\ket i.
\label{eq:Phi-state-vector}
\end{equation}
Here $\ket i$ are the bare atomic states featured in the definition
of $\hat{H}_{0}$ in Eq.~(\ref{eq:H0}). To account for the interaction between the atoms, we use the Gross--Pitaevskii approach for the multicomponent BECs \citep{PitaevskiiStringari}. In practice, this amounts to supplementing the Schrödinger equations for $\Phi_{i}(\boldsymbol{r})$ with the nonlinear terms, thereby promoting the single-particle wave functions to the wave functions (order parameters) of the components of the condensate \citep{Juzeliunas2005}. This way we arrive at

\begin{align}
\i\hbar\dot{\Phi}_{1} & =\left(-\frac{\hbar^{2}}{2m}\Delta+V+g_{11}|\Phi_{1}|^{2}+g_{12}|\Phi_{2}|^{2}\right)\Phi_{1}+\hbar\Omega_{1}^{*}\Phi_{3},\nonumber \\
\i\hbar\dot{\Phi}_{2} & =\left(-\frac{\hbar^{2}}{2m}\Delta+V+g_{12}|\Phi_{1}|^{2}+g_{22}|\Phi_{2}|^{2}\right)\Phi_{2}+\hbar\Omega_{2}^{*}\Phi_{3},\nonumber \\
\i\hbar\dot{\Phi}_{3} & =\left(-\frac{\hbar^{2}}{2m}\Delta+V-\varDelta-\i\frac{\Gamma}{2}\right)\Phi_{3}+\hbar\Omega_{1}\Phi_{1}+\hbar\Omega_{2}\Phi_{2}.\label{eq:full-gpe}
\end{align}

For atoms adiabatically following the dark state, the population of the excited state $\ket 3$ described by the wave function $\Phi_{3}$ is small at all times,
so atom collisions are not taken into account in the equation for
$\Phi_{3}$. The coefficients $g_{ij}$ describing interaction between the atoms in the corresponding internal states are related to the $s$-wave
scattering lengths $a_{ij}$ as $g_{ij}=4\pi\hbar^{2} a_{ij}/m$. The multicomponent
wave function of the BEC is normalized to the total number of atoms,
$\sum_{i=1}^{3}\int\d\boldsymbol{r}\,|\Phi_{i}|^{2}=N$. We will also
use parameters $\eta_{i}$ to quantify the relative population of
each component:
\begin{equation}
\eta_{i}=\frac{1}{N}\int|\Phi_{i}|^{2}\,\d\boldsymbol{r}.
\end{equation}

To obtain an equation for the evolution
of atoms adiabatically following the dark state, we introduce the
following superpositions of $\Phi_{1}$ and $\Phi_{2}$ describing
the wave functions for atoms in the dark and bright states introduced
in the previous Section \ref{subsec:Hamiltonian}: 
\begin{equation}
\begin{split}\Phi_{{\rm D}} & =\frac{1}{\sqrt{1+|\zeta|^{2}}}(\Phi_{1}-\zeta^{*}\Phi_{2}),\\
\Phi_{{\rm B}} & =\frac{1}{\sqrt{1+|\zeta|^{2}}}(\zeta\Phi_{1}+\Phi_{2}).
\end{split}
\label{eq:PhiD-PhiB}
\end{equation}
Inserting these definitions into Eq.~(\ref{eq:full-gpe}), we obtain
a set of equations for $\Phi_{{\rm D}}$, $\Phi_{{\rm B}}$ and $\Phi_{3}$,
in which the dark-state and bright-state wave functions $\Phi_{{\rm D}}$ and
$\Phi_{{\rm B}}$ are coupled via non-adiabatic terms. Under the adiabatic
assumption of slow center-of-mass motion for atoms in the dark internal states,
justified for the systems to be considered, and the assumption of
equal nonlinear couplings ($g_{11}=g_{12}=g_{22}=\beta$), we reach
the dark-state GPE \citep{Juzeliunas2005}

\begin{equation}
\i\hbar\dot{\Phi}_{{\rm D}}=\frac{1}{2m}(-\i\hbar\nabla-\boldsymbol{A})^{2}\Phi_{{\rm D}}+(U+V)\Phi_{{\rm D}}+\beta|\Phi_{{\rm D}}|^{2}\Phi_{{\rm D}}.\label{eq:ds-gpe}
\end{equation}
It is worth noting that, although $\Phi_{\rm D}$ is the solution to a nonlinear mean-field equation for a BEC, the term “dark-state wave function” remains appropriate. This state contains no atoms in the internal state $\ket{3}$, and, provided the adiabatic approximation holds, this remains true at all times, since the equation for $\Phi_{\rm D}$ is fully decoupled from those for $\Phi_{\rm B}$ and $\Phi_3$. In contrast, the equation for the “bright-state BEC” $\Phi_{\rm B}$ is coupled to $\Phi_3$, allowing atoms in $\Phi_{\rm B}$ to transition into $\ket{3}$ and decay.

It is apparent from Eq.~(\ref{eq:ds-gpe}) that reduction to the dark-state manifold results in the appearance
of geometric vector and scalar potentials $\boldsymbol{A}(\boldsymbol{r})$
and $U(\boldsymbol{r})$ emerging due to the position-dependence of
the atom--light coupling and given by \citep{Juzeliunas2005EIT,Juzeliunas2005}
\begin{equation}
\begin{split}\boldsymbol{A} & =\i\hbar\frac{\zeta^{*}\nabla\zeta-\zeta\nabla\zeta^{*}}{2(1+|\zeta|^{2})},\\
U & =\frac{\hbar^{2}}{2m}\frac{\nabla\zeta^{*}\cdot\nabla\zeta}{(1+|\zeta|^{2})^{2}}.
\end{split}
\end{equation}
For the beams given by Eq.~(\ref{eq:beams}) the geometric gauge
potentials do not depend on $z$ and read

\begin{equation}
\begin{split}\boldsymbol{A} & =-\frac{\hbar\nu}{a}\frac{\left(\frac{\rho}{a}\right)^{2\nu-1}}{1+\left(\frac{\rho}{a}\right)^{2\nu}}\boldsymbol{e}_{\varphi},\\
U & =\frac{1}{m}\left(\frac{\hbar\nu}{a}\right)^{2}\frac{\left(\frac{\rho}{a}\right)^{2\nu-2}}{\left[1+\left(\frac{\rho}{a}\right)^{2\nu}\right]^{2}}.
\end{split}
\label{eq:A_U}
\end{equation}
Here $\boldsymbol{e}_{\varphi}$ denotes a unit vector in the azimuthal
direction. The artificial magnetic field resulting from the geometric
potential $\boldsymbol{A}$ of Eq.~(\ref{eq:A_U}) is 
\begin{equation}
\boldsymbol{B}=\nabla\times\boldsymbol{A}=-\frac{2m}{\hbar}U\boldsymbol{e}_{z}\,,\label{eq:B}
\end{equation}
 where $\boldsymbol{e}_{z}$ is a unit vector directed along the
$z$-axis. For the light beams given by Eq.~(\ref{eq:beams}), the effective magnetic field is thus proportional to the scalar potential.

Once the solution of Eq.~(\ref{eq:ds-gpe}) is found, one can return
to the wave functions $\Phi_{1}$ and $\Phi_{2}$ of the internal-state
basis according to 

\begin{equation}
\begin{split}\Phi_{1} & =\frac{1}{\sqrt{1+|\zeta|^{2}}}\Phi_{{\rm D}},\\
\Phi_{2} & =-\frac{\zeta}{\sqrt{1+|\zeta|^{2}}}\Phi_{{\rm D}}.
\end{split}
\label{eq:ds-comps}
\end{equation}
This follows from Eq.~(\ref{eq:PhiD-PhiB}) with $\Phi_{{\rm B}}=0$,
since for adiabatic atomic motion in the dark state described by Eq.~(\ref{eq:ds-gpe})
for the wave-function $\Phi_{{\rm D}}$, the system is almost completely
in the dark state with a negligible population of the bright and excited
states.

In this way, when the atoms forming the BEC are in the dark state, the components $\Phi_{1}$ and $\Phi_{2}$ are related to each other via Eq.~\eqref{eq:ds-comps}, in which the component $\Phi_{2}$ has an extra factor $\zeta=\Omega_1/\Omega_2$. This factor is proportional to $\e^{\i\nu\varphi}$, as the Rabi frequency $\Omega_1$ has a vortex with a winding number $\nu$. As a result, at least one of the components $\Phi_{1}$ or $\Phi_{2}$ should have a vortex when the BEC atoms are in the dark state.

Our analysis will be mostly based on numerical calculations of the
stationary states described by the full three-component wave function,
governed by Eqs.~(\ref{eq:full-gpe}). We will concentrate on the
solutions belonging to the dark-state manifold, referring to the centre
of mass state for the dark state atoms with the lowest energy as the
ground state. Moreover, we will be interested in stationary vortex
solutions of the form $\Phi_{i}(\boldsymbol{r},t)\propto f_{i}(\rho)\e^{\i l_{i}\varphi}\e^{-\i\mu t/\hbar}$,
where $i=1,2,3$ and $\mu$ is the chemical potential of the system,
and we will refer to the integers $l_{i}$ as vorticities for each
internal state. On the other hand, Eq.~(\ref{eq:ds-gpe}) for the
dark state, valid under the assumption of sufficiently slow atomic
motion and equal intracomponent scattering lengths, will be used to
perform an analytic consideration in Section \ref{subsec:Analysis-based-on-ds-GPE}.

We remark that the dark state potentials arising from coupling BECs with LG beams have previously been considered (see, in particular, Ref.~\citep{Juzeliunas2005}). However, the stationary states emerging under continuous interaction with the laser beams have not been analyzed before.

\subsection{Trap geometry and parameter values\label{subsec:Trap-geometry-and}}

As the geometric vector and scalar potentials $\boldsymbol{A}$ and $U$ of Eq.~(\ref{eq:A_U}),
corresponding to the light beams (\ref{eq:beams}), do not depend on
$z$, we can consider the two-dimensional (2D) motion of dark-state atoms in the
$xy$ plane for fixed $z$. Specifically, for a cylindrically symmetric
setup of interest, in the dark-state GPE (\ref{eq:ds-gpe}) we can
separate the variables as $\Phi_{{\rm D}}(\rho,\varphi,z,t)=\Phi_{{\rm D}}^{(2)}(\rho,\varphi,t)Z(z)$
and take $Z(z)=1/\sqrt{d}$, thereby using the Thomas--Fermi approximation
for the atomic motion in the $z$-direction extended over the distance
$d$. The resulting equation for $\Phi_{{\rm D}}^{(2)}(\rho,\varphi,t)$
has the same form as Eq.~(\ref{eq:ds-gpe}) but the interaction strength
changes as $\beta\to\beta/d$. A similar procedure can be performed
for the exact system (\ref{eq:full-gpe}), yielding accurate results
in the adiabatic regime.

In our numerical simulations we consider a BEC cloud of $N=5\times10^{4}$
atoms confined in a cylindrical trap \citep{Gaunt2013} of the radius
$R$ and the extent $d$ along the $z$-axis with $R=d=\unit[15.0]{\mu m}$.
The states $\ket 1$ and $\ket 2$ are taken to be two hyperfine levels
of $^{87}{\rm Rb}$: $\ket 1=\ket{F=1,m_{F}=-1}$ and $\ket 2=\ket{F=2,m_{F}=1}$,
while state $\ket 3$ is the $5\,{}^{2}{\rm P}_{\frac{3}{2}}$ state,
whose decay rate is $\Gamma=\unit[38.1]{MHz}$ \citep{Gutterres2013}.
The scattering lengths are \citep{Egorov2013}: $a_{11}=100a_{0}$,
$a_{12}=98.0a_{0}$, $a_{22}=95.4a_{0}$, where $a_{0}$ is the Bohr
radius. The assumption of equal scattering lengths, used in deriving
the dark-state GPE, is thus justified for this system.

The typical values of the Rabi frequencies used in $\Lambda$-scheme
experiments range from $\sim\unit[10]{MHz}$ to $\sim\unit[1]{GHz}$.
Our calculations show that the dark-state GPE yields very accurate
results [compared to the solution of the full system of GPEs (\ref{eq:full-gpe})] already at $\Omega_{0}=\unit[1]{MHz}$,
meaning that the lowest states of the dark-state manifold indeed become
effectively decoupled from the states of the $\ket{\pm}$ manifolds.
Generally, the non-adiabatic coupling may be disregarded provided $U/\hbar\ll\Omega(\boldsymbol{r})$ for all $\boldsymbol{r}$  \citep{Juzeliunas2005,Juzeliunas2005EIT,Lacki2016},
where $\Omega\equiv \Omega(\boldsymbol{r})$ 
is the total Rabi frequency given by Eq.~\eqref{eq:pm-eigenvalue_eq}.
For $\nu=1$, the maximum of $U$ is located at the origin, while
$\Omega$ attains its minimal value there (we neglect the beam-waist
term $\e^{-\rho^{2}/w_{0}^{2}}$ in the present reasoning), equal
to $\Omega(\boldsymbol{\boldsymbol{r}}=0)=\Omega_{0}$. However, even
for the lowest value of $a$ considered in this work ($a=\unit[0.3]{\mu m}$),
one finds $U/\hbar=\unit[8]{kHz}$, which is orders of magnitude smaller
than the typical value of $\Omega_{0}$. Furthermore, although generally
including the decay rate of $\ket 3$ improves the validity of the
dark-state description \citep{Lacki2016,Jendrzejewski2016,Gvozdiovas2023},
in the regime in question we obtained equally accurate results even
for $\Gamma=0$ --- again because $\Omega_{0}$ is very large compared
to $U$. Finally, we mention the effect of the detuning $\varDelta$.
Non-zero positive values of $\varDelta$ shift the manifold of the $\ket -$ states down in
energy, further reducing the coupling with the states of the $\ket{{\rm D}}$
manifold. On the other hand, the states of the $\ket +$ manifold
are then also shifted to lower energies, and this enhances the coupling
with the $\ket{{\rm D}}$-manifold states. Thus, the dark-state GPE
is expected to be valid for some intermediate values of $\varDelta$.
For example, we have found that for $\Omega_{0}\sim\unit[10]{MHz}$,
values of $\varDelta$ from 0 to up to $\sim10\Omega_{0}/a^\nu$ lead to
good agreement between the dark-state analysis and the full treatment.
Meanwhile, negative values of $\varDelta$ shift to higher energies the manifold of  $\ket -$ states. This facilitates coupling between the $\ket -$ and $\ket{{\rm D}}$ states, reducing the accuracy of the dark-state description.

In all our calculations we used $\Omega_{0}=\unit[20.0]{MHz}$
and $\varDelta=10\Omega_{0}/a^{\nu}$, where the parameter $a$ characterizes the Rabi frequency of the vortex light beam in Eq.~\eqref{eq:beams}. The beam-waist term $\e^{-\rho^{2}/w_{0}^{2}}$
can be set to unity assuming the waist is much larger than the radial
extent of the cloud. Moreover, the term $\e^{-\rho^{2}/w_{0}^{2}}$ plays no role in the adiabatic dynamics of atoms in
the dark state, since it drops out of the relative Rabi frequency $\zeta$ given by Eq.~(\ref{eq:zeta}).
However, we did include this term in the exact numerical calculations by setting $w_{0}=2R$ --- while
the results were the same, this provided a slight speed-up for the
calculations. 

Henceforth we adopt the dimensionless units, measuring length in units
of $R$, energy in units of kinetic energy $E_{R}=\hbar^{2}/2mR^{2}$
corresponding to the momentum $k_{R}=1/R$, and time in units of $\hbar/E_{R}$.
In these units the radial coordinate $\rho$ varies from 0 to 1. We
refer the reader to the Appendix for additional details on the units
and the dimensionless form of the equations used for numerical calculations.

\begin{figure*}
\begin{centering}
\includegraphics{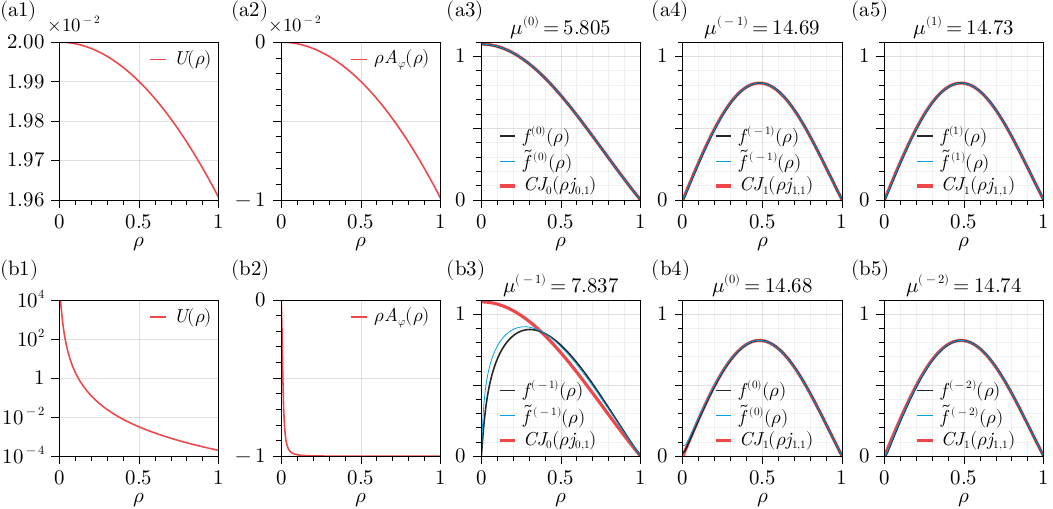}
\par\end{centering}
\caption{\label{fig:beta0}The potentials $U$, $\rho A_{\varphi}$ and the
solutions of Eq.~(\ref{eq:bessel}). Panels (a1)--(a5) show,
respectively, the curves $U$, $\rho A_{\varphi}$, and three solutions
of Eq.~(\ref{eq:bessel}) (with $\beta=0$) for $\nu=1$ and $a=10$. In panels (a3)--(a5), curves $\tilde{f}^{(l)}(\rho)$ depict the solution obtained under an additional assumption $U=0$. Panels (b1)--(b5)
show the same as panels (a1)--(a5) for $\nu=1$ and $a=0.01$.
The depicted wave functions are normalized such that $\protect\int_{S}|\psi^{(l)}(\rho,\varphi)|^{2}\protect\d S=2\pi\protect\int_{0}^{1}[f^{(l)}(\rho)]^{2}\rho\,\protect\d\rho=1$.}
\end{figure*}

\section{Stationary vortex solutions}

\subsection{Approximate analysis of motion of dark-state atoms\label{subsec:Analysis-based-on-ds-GPE}}

To gain physical insight into the expected structure of the possible
stationary states of the system undergoing the 2D
motion, it is instructive to consider the dark-state GPE (\ref{eq:ds-gpe})
first. To find stationary solutions, we put $\Phi_{{\rm D}}(\boldsymbol{r},t)=\e^{-\i\mu t}\psi(\boldsymbol{r})$
and further separate the variables due to the cylindrical symmetry
of the system:
\begin{equation}
\psi(\rho,\varphi)=f(\rho)\e^{\i l\varphi}\label{eq:psi}
\end{equation}
 with $f(\rho)$ being real and $l$ being integer. In the specific case of the geometric potentials given in Eq.~(\ref{eq:A_U}), this leads to
the equation of the Bessel type (with an additional nonlinear term) for $f\equiv f(\rho)$:
\begin{equation}
\rho^{2}\partial_{\rho}^{2}f+\rho\partial_{\rho}f+[(\mu-U-\beta f^{2})\rho^{2}-(l-\rho A_{\varphi})^{2}]f=0.\,\label{eq:bessel}
\end{equation}
Here the external trapping potential term is not written explicitly, but it controls
the boundary conditions: $f(1)=0$.

To allow for an analytical treatment, in the following Sections \ref{subsubsec:large-a} and \ref{subsubsec:small-a} we will neglect the interaction term by setting $\beta=0$. Our numerical calculations
confirm that the conclusions obtained regarding the value of $l$
of the ground-state wave function remain valid in the presence of
interactions. We will analyze two limiting regimes for the solutions of Eq.~(\ref{eq:bessel}) with $\beta=0$. The regimes are defined by the distance $a$ describing the extent of the geometric scalar potential $U$ and the effective magnetic field $\boldsymbol{B}$ given by Eqs.~\eqref{eq:A_U}--\eqref{eq:B}.

\subsubsection{Large $a$\label{subsubsec:large-a}}

First, we consider the case when the distance $a$ is large compared to the system radius $R$, so that $\rho\ll a$.
Then, according to Eq.~(\ref{eq:A_U}), for $l\ne0$ one can
neglect the term $\rho A_{\varphi}\approx-\left(\rho/a\right)^{2\nu}\nu$ in comparison with $l$ in Eq.~\eqref{eq:bessel}, whereas for $l=0$ this term may be disregarded because $A_{\varphi}^2\ll\mu$. 
In a similar way, the geometric scalar potential $U$ is
small in comparison with $\mu$ and thus can be also neglected.
In such a situation, the solutions satisfying the boundary condition are the properly scaled
Bessel functions of the first kind $J_{l}$. From the sequence of
zeros of $J_{l}$ we find that the ground state corresponds to $l=0$,
and the degenerate pair of the lowest excited states corresponds to
$l=\pm1$. Explicitly, the chemical potential $\mu^{(l)}$ (the eigenenergy) of the
ground state is given by $\mu^{(0)}=(j_{0,1})^{2}$, where $j_{l,n}$
is the $n$th root of the Bessel function $J_{l}(x)$. The corresponding
wave function $\psi^{(l)}$ is given by $\psi^{(0)}(\rho,\varphi)=f^{(0)}(\rho)=CJ_{0}(\rho j_{0,1})$, with $C$ being a normalization constant. 

These findings are illustrated
in Figs.~\ref{fig:beta0}(a1)--(a5), corresponding to $a=10$, $\nu=1$.
Figures \ref{fig:beta0}(a1) and \ref{fig:beta0}(a2) display the
curves of $U(\rho)$ and $\rho A_{\varphi}(\rho)$. It is apparent
that the two functions are low in magnitude compared to the chemical potential and can be disregarded
as noted above. Indeed, numerically solving Eq.~(\ref{eq:bessel})
(with $U$ and $\rho A_{\varphi}$ included) we find that the chemical
potential of the ground state is $\mu^{(0)}=5.805$, close to the
limiting value $(j_{0,1})^{2}\approx5.783$. The solution $f^{(0)}(\rho)$
also matches the limiting solution $J_{0}(\rho j_{0,1})$, as shown
in Fig.~\ref{fig:beta0}(a3). The blue curve additionally shows
the solution $\tilde{f}^{(0)}(\rho)$ obtained from Eq.~(\ref{eq:bessel})
with $U$ disregarded, confirming that the scalar potential has almost
no impact on the solution. Finally, Figs.~\ref{fig:beta0}(a4)--(a5)
display the excited states corresponding to $l=\pm1$; the degeneracy
is slightly lifted by the $\rho A_{\varphi}$ term.

To draw conclusions regarding the motion of atoms, we have to return to the
wave functions of the two components given in Eq.~(\ref{eq:ds-comps}).
Substituting the ground state solution $\Phi_{{\rm D}}^{(0)}=\e^{-\i\mu t}\psi^{(0)}(\boldsymbol{r})$,
we find that in the ground state, $\Phi_{1}^{(0)}$ is vortex-free
and is localized near the origin (with a maximum at $\rho=0$), while
$\Phi_{2}^{(0)}$ describes a vortex having zero density at the origin
and rotating around the first component with vorticity equal to $l_{2}=l+\nu=+1$.
It should be noted that, according to Eq.~(\ref{eq:ds-comps}), the
vorticities of the components always differ by $\nu$ units. Therefore,
at least one of the two components will necessarily be in a vortex
state as long as the spatial part of $\Phi_{{\rm D}}$ is of the form
(\ref{eq:psi}). For example, even if $l=0$, then $\Phi_{1}$ will
be vortex-free, while $\Phi_{2}$ will have vorticity equal to $\nu$.

\subsubsection{Small $a$\label{subsubsec:small-a}}

Now let us consider the case when the distance $a$ is small compared to the system radius, $a\ll R$.
In that case, one has $\rho\gg a$
for most of the area occupied by the condensate, except for a small area close to the center, $\rho \lessapprox a$ in which the geometric scalar and vector potentials are concentrated. For $\rho\gg a$, it follows from Eq.~(\ref{eq:A_U}) that $\rho A_{\varphi}(\rho)\to-\nu$. The ground-state solution
can be found by choosing $l$ such that $l+\nu=0$; the solution
will again be given by $J_{0}$. Thus, for $\rho\gg a$ the ground
state wave function is given by $\psi^{(-\nu)}(\rho,\varphi)=CJ_{0}(\rho j_{0,1})\e^{-\i\nu\varphi}$.
However, close to the origin, where $\rho \lessapprox a$, the
radial part of the solution will have a different form. Indeed, $l$ is now fixed
to $-\nu$, and we can neglect the term $\rho A_{\varphi}$. The
solution in the region $\rho\ll a$ is then given by $J_{\nu}$,
having a zero at the origin as a vortex solution should. Meanwhile,
in the region $\rho\gg a$, the two lowest excited states will come
in a degenerate pair having the radial dependence $J_{1}$ and corresponding
to $l+\nu=\pm1$. Close to the origin ($\rho\ll a$), they will have
the radial dependencies $J_{\pm1-\nu}$. These results are illustrated
in Figs.~\ref{fig:beta0}(b1)--(b5) for the case where $a=0.01$, $\nu=1$.
In Fig. \ref{fig:beta0}(b1), it is apparent that the maximum value of $U$ greatly exceeds
that of the chemical potential. However, the presence of the potential does not appreciably influence the resulting wave functions due to the potential being localized at the origin. Figure \ref{fig:beta0}(b2)
displays the curve $\rho A_{\varphi}$, which has a value close to
$\nu=-1$ in most of space occupied by the condensate. In agreement
with the preceding arguments, the state with the lowest chemical potential
is obtained for $l=-1$ {[}see Fig.~\ref{fig:beta0}(b3){]}. Away
from the origin, the wave function tends to $J_{0}$, while at the
origin it attains a zero. Further reducing $a$ increases the steepness
of the wave function near the origin and enhances the match with $J_{0}$
(not shown). The degenerate pair of the lowest excited states is shown
in Figs.~\ref{fig:beta0}(b4)--(b5). These wave functions mostly
follow the shape of $J_{1}$ except for the small portion of space near
the origin where the condition $\rho\gg a$ no longer holds. There, the state with $l=+1-\nu=0$ has the form of a $J_{0}$ function, while the $l=-1-\nu=-2$
state has the shape of $J_{2}$.

Thus, if parameter $a$ is small enough so that the condition $\rho\gg a$
holds in most of space occupied by the condensate, the 
dark-state solution of the lowest energy will be the one corresponding to $l=-\nu$. In
this limit the gradient of the phase $\phi$ of wave function equals
$-\boldsymbol{e}_{\varphi}\nu/\rho$, thereby canceling the contribution of vector potential
$\boldsymbol{A}$ to the velocity field: $\boldsymbol{v}\propto\nabla\phi-\boldsymbol{A}\to0$.
As a result, $\Phi_{1}$ will be in a vortex state, while $\Phi_{2}$
will be vortex-free.

\subsection{Full treatment}

\begin{figure*}
\begin{centering}
\includegraphics{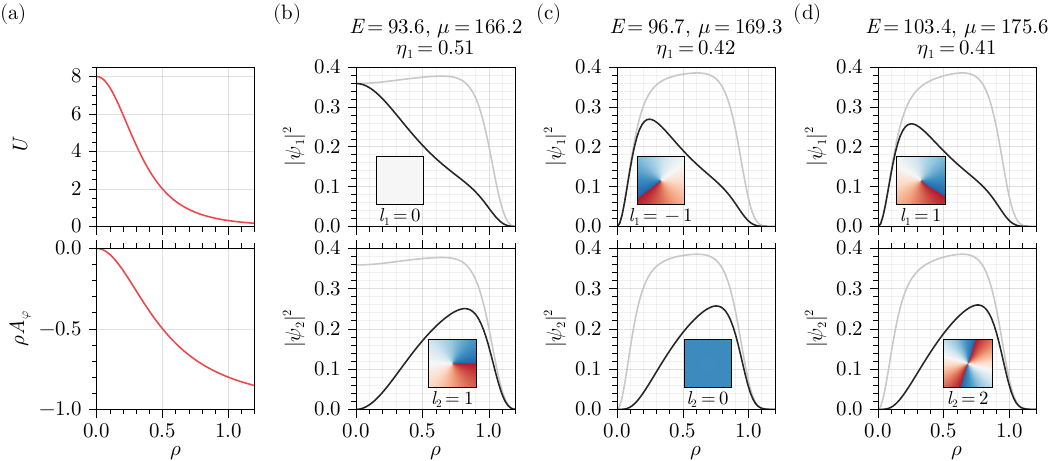}
\par\end{centering}
\caption{\label{fig:a0.5}The potentials $U$ and $\rho A_{\varphi}$, as well as the
solutions of Eq.~(\ref{eq:full-gpe}) for $\nu=1$ and $a=0.5$. (a)
Potentials $U$ and $\rho A_{\varphi}$. (b), (c), and (d) show, respectively,
three lowest-energy solutions of Eq.~(\ref{eq:full-gpe}); wave function normalization is $\sum_{i=1}^{3}\int_S\,|\psi_{i}|^{2}\,\d S=1$.
The black lines in the upper and lower panels show, respectively, the radial cuts of the densities $|\psi_{1}|^{2}$ and $|\psi_{2}|^{2}$ of the first and the second component. The gray lines show the total density.
The values of the density of the third component are vanishingly small in all cases (calculation yields values no larger than $10^{-7}$) and are therefore not shown. For the same reason,
the relative occupation of the second component can be taken to be
$\eta_{2}=1-\eta_{1}$. The vorticities $l_{i}$ of the two components
are displayed together with the insets showing the two-dimensional
phase profiles in the $xy$-plane. The value of the phase is color-coded
as follows: $\text{dark blue}=-\pi$, $\text{white}=0$, $\text{dark red}=\pi$.}
\end{figure*}

\begin{figure*}
\centering{}\includegraphics{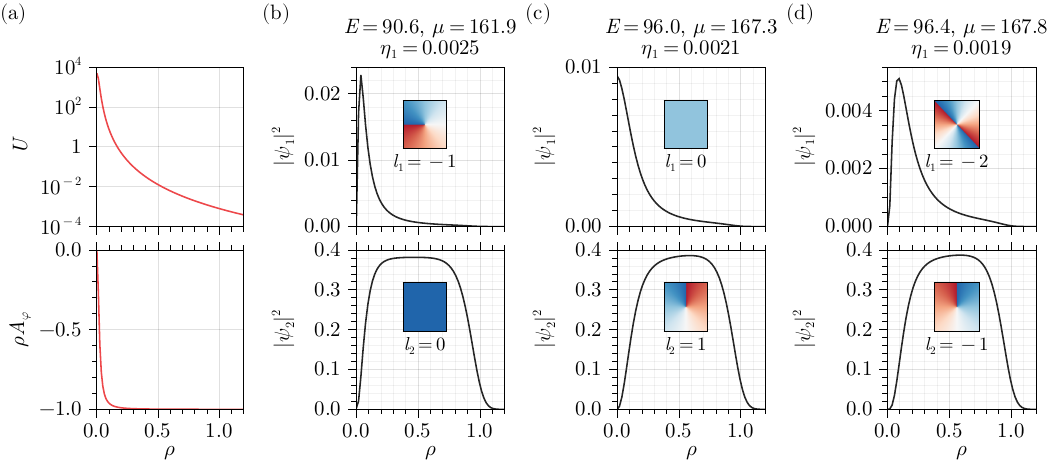}\caption{\label{fig:a0.02}Same as Fig.~\ref{fig:a0.5} for $\nu=1$ and $a=0.02$.
The total densities are not explicitly shown since they would appear to largely coincide with $|\psi_{2}|^{2}$ due to the highly imbalanced occupation of the two modes.
}
\end{figure*}

\begin{figure*}
\centering{}\includegraphics{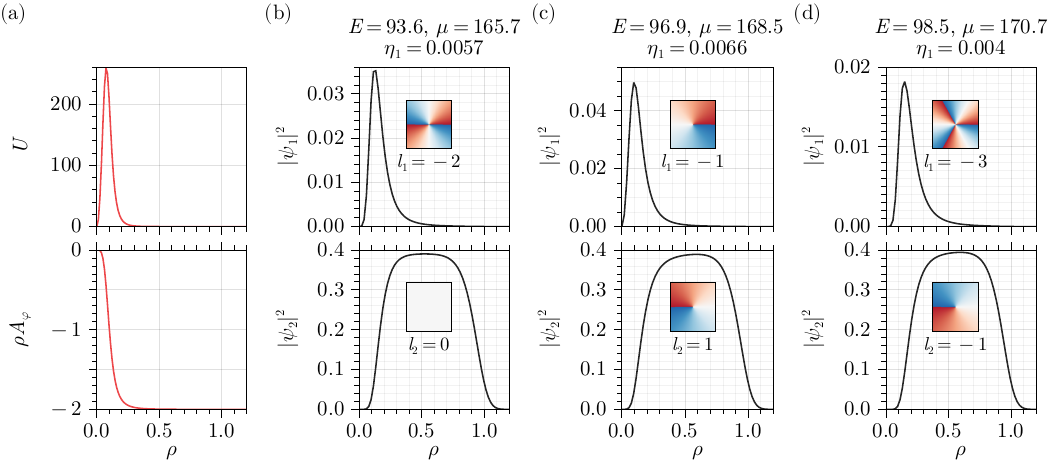}\caption{\label{fig:L2_a0.1}Same as Fig.~\ref{fig:a0.02} for $\nu=2$ and $a=0.1$.}
\end{figure*}

We will now turn to the solutions of the full coupled GPEs (\ref{eq:full-gpe}), including the atom--atom interactions,
and study the structure of the lowest-energy states of the dark-state
manifold. To obtain the numerical solutions, we employed the method
of imaginary-time evolution: one makes a change $t\to-\i\tau$ and
propagates the equations in time $\tau$ until convergence is reached,
starting from a certain trial wave function. The algorithm will then
converge to the lowest-energy state provided it has nonzero overlap
with the trial wave function. To enable convergence to the states
of the dark-state manifold, we used the solution of Eq.~(\ref{eq:ds-gpe})
as the trial function for solving coupled GPEs (\ref{eq:full-gpe}). In turn, the trial wave functions required to solve Eq.~(\ref{eq:ds-gpe}) were constructed by solving this equation using the Thomas–Fermi approximation and multiplying the solution by a phase factor $\e^{\i l\varphi}$ with a chosen value of $l$. Since states of the form (\ref{eq:psi}) with different $l$ are orthogonal, the trial wave function of this form converges to a state with the same value of $l$. This allowed us to obtain stationary solutions of different vorticities. The ground state can then be found by ordering the states based on the
value of the energy per particle $E$.
The population of each component was calculated from the obtained solutions rather than being enforced.
We refer the reader to the Appendix for additional numerical details.

In a stationary state, the wave functions of the components have the
form $\Phi_{i}(\boldsymbol{r},t)=\e^{-\i\mu t}\psi_{i}(\boldsymbol{r})$.
Below, we show the results obtained using full three-level calculation
so as to take into account the different scattering lengths. If they
are taken to be equal, then the results match with the dark-state
calculations based on solving Eq.~(\ref{eq:ds-gpe}) or Eq.~(\ref{eq:bessel}) (the resulting
total chemical potentials agree with at least three-digit accuracy).

We start by setting $\nu=1$ and $a=0.5$. Such a value of $a$ corresponds
to an intermediate regime whereby the term $\rho A_{\varphi}$ is
not small (compared to unity) and cannot be approximated by $-\nu$, as
one can see in the lower panel of Fig.~\ref{fig:a0.5}(a). Numerical calculations
show that ground state is the one corresponding to an $l=0$ solution
of the dark-state GPE (\ref{eq:ds-gpe}). As noted above, this corresponds
to $\Phi_{1}$ being vortex-free and $\Phi_{2}$ being in a vortex
state. Such a solution is shown in Fig.~\ref{fig:a0.5}(b). It is
apparent that the first component fills the core of the second component,
a situation known to stabilize vortices with vorticities higher than $l=1$ \citep{Patrick2023,Richaud2023}. Notably, which of the components acquires
vorticity is determined by the inequivalent coupling fields $\Omega_{1}$
and $\Omega_{2}$ rather than the intrinsic properties of the two
states (the scattering lengths).
The total density of the BEC, shown in gray in Fig.~\ref{fig:a0.5}(b), indicates that the condensate as a whole does not have a pronounced density dip at the origin, meaning that the core of the vortex is ``completely'' filled.

Two lowest excited states are shown in Figs.~\ref{fig:a0.5}(c)--(d).
Their energies per particle are 96.7 and 103.4 units, respectively, showing that
the term $\rho A_{\varphi}$ has lifted the degeneracy. Notably, the
``interaction potentials'' $g_{ij}|\Phi_{k}|^{2}$ do not depend
on the vorticities of the components and, therefore, do not lift the
degeneracy. In the first excited state, the first component is in
a vortex state, rotating in the direction opposite to the direction of the vector field $\boldsymbol{A}$, while the second component is vortex-free.
The vortex on the first component mostly occupies the
space at the center of the second component. Such an approximate phase
separation is in line with the criterion $g_{12}^{2}>g_{11}g_{22}$
\citep{Ao1998,Timmermans1998} which, however, does not take into
account the kinetic energy \citep{Wen2012}, which is especially important
in the present setting. In the second excited state, vorticities of the two components are given by $l_1=1$ and $l_2=2$, while the spatial profiles are almost the same as those of the respective components of the first excited state. 
In both cases, the components contribute to a nearly uniform total density of the cloud in the bulk.

Let us now study the regime where the distance $a$ is much smaller that the system radius $R$, specifically $a=0.02$. In this case the condition
$\rho\gg a$ holds in most of the space, and $\rho A_{\varphi}$ tends
to $-\nu$ already in the vicinity of the origin {[}see Fig.~\ref{fig:a0.02}(a),
lower panel{]}. The ground state now corresponds to the
$l=-\nu=-1$ solution of the dark-state GPE (\ref{eq:ds-gpe}). This means that $\Phi_{1}$ is a vortex state, and it occupies
a small region of space near the origin [see Fig.~\ref{fig:a0.02}(b)]. Specifically, the density of the first component is concentrated in a region of the order of the healing length $\xi_2$ of the second component. Assuming all atoms are in the second component, we have (in dimensionless units) $\xi_2 = \sqrt{S/g_{22}}=0.086$, where we approximately took $S=\pi$ for the occupied area.
The localized nature of the vortex results from the $\frac{1}{\sqrt{1+|\zeta|^{2}}}$
term in the expression for $\Phi_{1}$ (\ref{eq:ds-comps}), which
effectively cuts the vortex off when $\rho\gg a$. Another consequence
is the small resulting population of the first component, which in the
present case is $\eta_{1}=0.25\%$. Thus, decreasing the $a$
parameter localizes the vortex more tightly, but also reduces the
population of the first component. The second component occupies the
space surrounding the first component and has zero vorticity.

The first excited state, on the other hand, has the typical structure
whereby the non-rotating component fills the core of the rotating
one {[}see Fig.~\ref{fig:a0.02}(c){]}. The almost-degenerate partner
of this state is shown in Fig.~\ref{fig:a0.02}(d). These two states
correspond, respectively, to the $l=0$ and $l=-2$ solutions of the
dark-state GPE (\ref{eq:ds-gpe}). The degeneracy results from the
fact that the kinetic energy is the same, whether the cloud does not
rotate ($l=0$) or rotates in the direction opposite to the $\boldsymbol{A}$ field
but with twice the speed ($l=-2$).

Localized vortices with vorticities higher than one can be analogously
created by using beams with $\nu>1$. The results for the case $\nu=2$
are presented in Fig.~\ref{fig:L2_a0.1}. Choosing $a=0.1$ already
leads to the potential $\rho A_{\varphi}$ tending to $-\nu$ in
most of space {[}see Fig.~\ref{fig:L2_a0.1}(a){]}. This leads to
a ground state with the first component having vorticity $l_{1}=-2$
and the second being vortex-free: $l_{2}=l_{1}+\nu=0$. The vortex
has a density maximum farther from origin compared to the $l_{1}=-1$
vortex in Fig.~\ref{fig:a0.02}(b) due, in part, to the former having
the form of $J_{2}$ and thus rising less sharply compared to $J_{1}$.
The degree of localization can be increased by reducing $a$, but
the number of atoms in the vortex component will then decrease. The
lowest excited states {[}see Figs.~\ref{fig:L2_a0.1}(c)--(d){]}
also contain localized vortices in the first component.

It is worth noting that the density profiles of both components remain essentially unchanged as $g_{12}$ is reduced into the miscible regime, $g_{12}^2 < g_{11} g_{22}$, provided the coupling constants $g_{ij}$ are of similar magnitude. In particular, the densities shown in Figs.~\ref{fig:a0.02}(b) and \ref{fig:L2_a0.1}(b) are virtually unaffected. This invariance is expected, as the localized-vortex configuration is sustained by the external laser beams rather than interparticle interactions, as demonstrated in Section III A, where interactions are entirely neglected.

\begin{figure*}
\centering{}\includegraphics{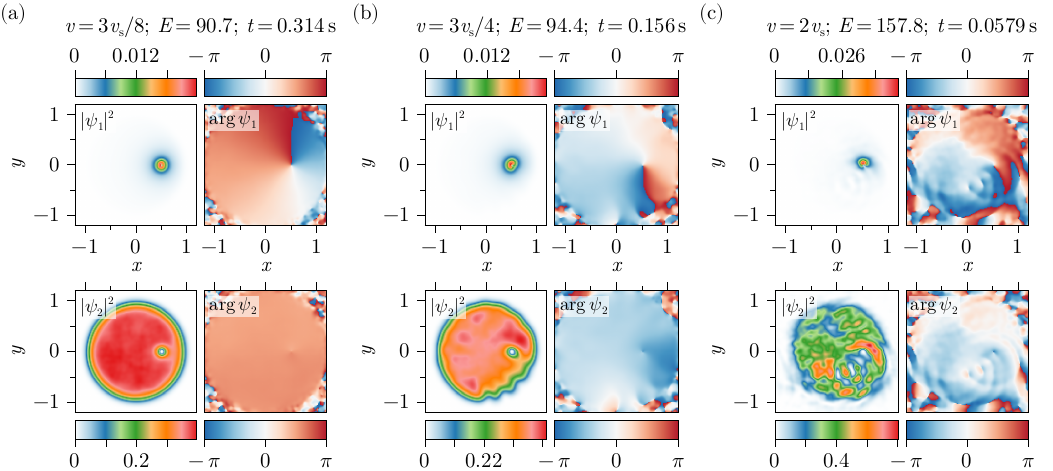}\caption{\label{fig:move}Wave functions of components 1 and 2 after completing
one circular sweep of the laser beams around the origin. The tangential
beam movement speed is (a) $v_{1}=3v_{\rm s}/8=\unit[150]{\mu m/s}$, (b) $v_{2}=3v_{\rm s}/4=\unit[300]{\mu m/s}$, and (c) $v_{3}=2v_{\rm s}=\unit[800]{\mu m/s}$. In all three cases the relative
population of the first component is $\eta_{1}\approx0.0025$, while the third component is essentially unpopulated.}
\end{figure*}

\section{Moving the vortex}

To demonstrate the stability of the localized lowest-energy vortex
states and show the high level of available control, we consider changing
the position of the vortex by moving the laser beams. A similar protocol
has been considered in Ref.~\citep{Di2023}, however, that work considered
moving a vortex created by a phase-imprinting technique. Our approach
is more robust because the laser beams create a pinned vortex which
cannot break free.

We focus on the regime $\nu=1$, $a=0.02$ studied in Fig.~\ref{fig:a0.02}
but this time consider an off-axis vortex created by centering the
laser beams at $(x,y)=(0.5,0)$. Once the state is prepared, the transverse profile of the laser beams $\Omega_1$ and $\Omega_2$ start
moving counterclockwise on a circular path around the origin at a constant
tangential speed $v$. Figure \ref{fig:move} shows the wave functions
obtained after completing one full circle. In Fig.~\ref{fig:move}(a)
corresponding to $v_{1}=\unit[150]{\mu m/s}$ (time to complete the
circle is $t=\unit[0.314]{s}$), it is apparent that the vortex of
the first component has retained its structure, and the density of
the second component has not been strongly distorted. The value of
$E$ has increased by only $3\%$ of the energy (per particle) gap
between the ground state and the first excited one. Here the velocity
is chosen to be lower than the speed of sound in the second component,
which equals to $v_{\rm s}=\hbar/(m\xi_2\sqrt{2})=\unit[400]{\mu m/s}$. Moving the vortex at twice
the speed, $v_{2}=2v_{1}$, results in setting the whole of the second
component in circular motion. The distortion of the density is apparent
in Fig.~\ref{fig:move}(b).

Finally, the case of supersonic movement is studied in Fig.~\ref{fig:move}(c)
corresponding to the movement speed $v_{3}=2v_{\rm s}$. The density profiles
of both components are strongly distorted and the energy is increased
1.5 times. Nevertheless, the first component retains its signature
phase profile, and no additional vortices are created in the
second component in the process.
Let us stress that in all three cases, the third component remained essentially unpopulated, and the total norm of the wave functions was conserved on the level of the numerical accuracy used, meaning that, in practice, no loss of atoms took place.

\begin{figure}
\centering{}\includegraphics{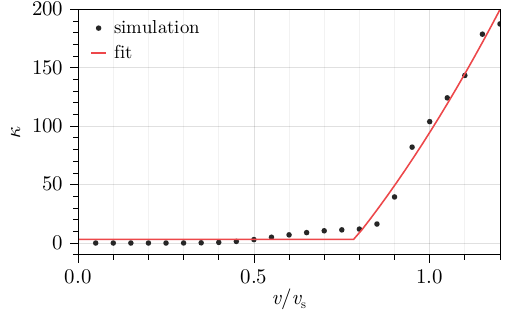}
\caption{
\label{fig:heating}
Condensate heating rate as a function of the movement speed of the beams.
}

\end{figure}

It is also instructive to use the localized vortex in the first component as a probe for the critical superfluid velocity $v_{\rm c}$ of the second component. In this case, the vortex acts as an impurity that can move without resistance provided its speed remains below $v_{\rm c}$. According to Landau’s criterion, the phononic excitation spectrum of the condensate sets the critical velocity equal to the sound velocity \cite{PethickSmith}. However, both experimental results \cite{Raman1999,Onofrio2000,Desbuquois2012,Weimer2015} and detailed theoretical studies \cite{Singh2016,Singh2017,Kiehn2022} indicate that $v_{\rm c}$ is typically lower, depending on factors such as impurity size \cite{Desbuquois2012}, the nature of the impurity potential (attractive or repulsive) \cite{Singh2016}, and the impurity trajectory (linear or circular) \cite{Singh2016,Kiehn2022}. To measure $v_{\rm c}$ in our system, we repeat the circular trajectory simulations, moving the beams for a fixed duration of $\unit[0.1]{s}$. We monitor the total energy per particle, $E(t)$, and fit it with a linear function to obtain the heating rate $\kappa = \mathrm{d}E/\mathrm{d}t$. Since the first component’s population is much smaller than that of the second, $E(t)$ primarily reflects the energy of the atoms in the second component. The results for velocities from $0.1v_{\rm s}$ to $1.2v_{\rm s}$ are shown in Fig.~\ref{fig:heating}, revealing a clear threshold behavior where $\kappa$ increases rapidly once the velocity exceeds a certain value.

We quantified this threshold by fitting the data to
\begin{equation}
\kappa(v)=\kappa_{0}+b\max(v^{2}-v_{{\rm c}}^{2},0),
\end{equation}
where $\kappa_{0}$, $b$, and $v_{\rm c}$ are fit parameters. This form is suitable for 2D systems \cite{Astrakharchik2004} and has proven effective in experimental analyses \cite{Desbuquois2012,Singh2017}. The least-squares fit shown in Fig.~\ref{fig:heating} yields $v_{\rm c} = 0.8v_{\rm s}$, which is lower than the sound velocity and consistent with the above observations. The precise dissipation mechanisms warrant further study and will be addressed in a future work.

\section{Conclusions}

In summary, we have studied the interaction of a two-component BEC
mixture with the light fields in a $\Lambda$-type configuration and
investigated the stationary states of the dark-state manifold. The
angular momentum of $\nu\hbar$ per photon carried by one of the
two fields leads to either one or both components being in a vortex
state, with their vorticities differing by $\nu$ units. In the regime
$a\ll R$, the structure of the stationary states is dictated
by the vector potential term. In this limit, $\boldsymbol{A}$ tends
to $-\frac{\nu}{\rho}\boldsymbol{e}_{\varphi}$ and its contribution
to the velocity field (and hence to the kinetic energy) can be canceled
if the dark state is described by a wave function $f(\rho)\e^{-\i\nu\varphi}$.
We have demonstrated that the lowest-energy state of the dark-state
manifold indeed has this form. In this case, the first component (the
one interacting with the LG beam) contains a vortex of vorticity $\nu$,
while the second component is vortex-free. The core of the former
vortex coincides with the center of the beams, while the density profile
of the vortex demonstrates a strong degree of localization as  the density falls off as $[1+(\rho/a)^{2}]^{-1/2}$ away
from the vortex core.
Such a vortex can be moved around by moving the laser beams. Provided
the movement speed is less than approximately half the speed of sound
in the second component, the shape of the vortex retains its structure during
the movement, and the density of the second component does not get
distorted.
More specifically, moving the vortex at a speed below the critical value of about $0.8v_{\rm s}$ causes almost no heating of the condensate, indicating its superfluid nature.

\begin{acknowledgments}
Y.\,B. would like to acknowledge the outstanding hospitality received
during his internship at the Okinawa Institute of Science and Technology
Graduate University, Quantum Systems Unit, where work on this project was initiated. This project was supported
by the Research Council of Lithuania (RCL) grant No.~S-LJB-24-2 and the JSPS Bilateral Program No.~JPJSBP120244202. This work has also received funding from COST Action POLYTOPO CA23134, supported by COST (European Cooperation in Science and Technology). 
\end{acknowledgments}

\section*{Appendix: Dimensionless units and numerical details}

\setcounter{equation}{0}
\renewcommand{\theequation}{A\arabic{equation}}

For our numerical calculations we used dimensionless units, measuring
length in units of cylindrical trap radius $R$ and energy in units of $E_{R}=\hbar^{2}/2mR^{2}$, and time in units of $\hbar/E_{R}$.
In these units, the system of equations (\ref{eq:full-gpe}) becomes

\begin{align}
\i\dot{\Phi}_{1} & =\left(-\Delta+V+g_{11}|\Phi_{1}|^{2}+g_{12}|\Phi_{2}|^{2}\right)\Phi_{1}+\Omega_{1}^{*}\Phi_{3},\nonumber \\
\i\dot{\Phi}_{2} & =\left(-\Delta+V+g_{12}|\Phi_{1}|^{2}+g_{22}|\Phi_{2}|^{2}\right)\Phi_{2}+\Omega_{2}^{*}\Phi_{3},\nonumber \\
\i\dot{\Phi}_{3} & =\left(-\Delta+V-\varDelta-\i\frac{\Gamma}{2}\right)\Phi_{3}+\Omega_{1}\Phi_{1}+\Omega_{2}\Phi_{2}.\label{eq:full-gpe-dimensionless}
\end{align}
Our calculations were carried out in the dark-state regime, where the amplitudes of $\Phi_3$ were vanishingly small. Consequently, the non-Hermitian nature of the dynamics arising from the imaginary decay term $\Gamma$ was effectively negligible, allowing us to adopt the conventional normalization used for Hermitian systems,
$\sum_{i=1}^{3}\int\d\boldsymbol{r}\,|\Phi_{i}|^{2}=1$
(the sum can equally be taken over $\Phi_1$ and $\Phi_2$ only). Furthermore, the energies and chemical potentials of the stationary states had vanishingly small imaginary parts, implying that the wave function norm remained essentially constant over time. We note that in our numerical calculations, the wave functions were normalized to unity rather than to the total particle number.

We take $R=\unit[15.0]{\mu m}$ and the mass of an $^{87}{\rm Rb}$
atom $m=\unit[1.443\times10^{-25}]{kg}$.
The amplitudes of the Rabi frequencies $\Omega_{1}$ and $\Omega_{2}$
are expressed in terms of $\Omega_{0}$ [see Eq.~(\ref{eq:beams})]
whose value is $\Omega_{0}=\frac{\unit[20.0]{MHz}}{E_{R}/\hbar}=1.23\times10^{7}$;
the decay rate is $\Gamma=\frac{\unit[38.1]{MHz}}{E_{R}/\hbar}=2.35\times10^{7}$,
and the detuning is $\varDelta=10\Omega_{0}/a^{\nu}$. The interaction
strengths are characterized by $g_{ij}=8\pi a_{ij}N/d$, where division
by the cylindrical trap height $d$ appears as a result of reduction
to the 2D equations (see main text). We take $d=R$, $N=5\times10^{4}$,
$a_{11}=100a_{0}$, $a_{12}=98.0a_{0}$, and $a_{22}=95.4a_{0}$,
where $a_{0}$ is the Bohr radius. This results in $g_{11}=443$,
$g_{12}=434$, $g_{22}=423$.

The radial profile of the cylindrical trap was modeled by a logistic
function \citep{Patrick2023}
\begin{equation}
V(\rho)=\frac{V_{0}}{1+V_{0}\e^{-b(\rho-\rho_{0})}}
\end{equation}
with $V_{0}=1000$, $b=17$, $\rho_{0}=0.7$, ensuring a steep rise
at $\rho\approx1$. The computational grid contained 129 points in
each dimension, spanning the interval $\rho\in[-1.2,1.2]$.

The dark-state GPE in dimensionless units becomes

\begin{equation}
\i\dot{\Phi}_{{\rm D}}=(-\i\nabla-\boldsymbol{A})^{2}\Phi_{{\rm D}}+(U+V)\Phi_{{\rm D}}+\beta|\Phi_{{\rm D}}|^{2}\Phi_{{\rm D}}\label{eq:ds-gpe-dimensionless}
\end{equation}
while the components of the artificial gauge field are given by 
\begin{equation}
\begin{split}\boldsymbol{A} & =-\frac{\nu}{a}\frac{\left(\frac{\rho}{a}\right)^{2\nu-1}}{1+\left(\frac{\rho}{a}\right)^{2\nu}}\boldsymbol{e}_{\varphi},\\
U & =2\left(\frac{\nu}{a}\right)^{2}\frac{\left(\frac{\rho}{a}\right)^{2\nu-2}}{[1+\left(\frac{\rho}{a}\right)^{2\nu}]^{2}}.
\end{split}
\end{equation}
We remind that the dark-state GPE is obtained under the assumption
of equal inter- and intracomponent scattering lengths; in our calculations
we used $\beta=g_{11}=443$.

The solutions $\Phi_{{\rm D}}$ of the dark-state GPE (\ref{eq:ds-gpe-dimensionless})
were obtained using the imaginary-time evolution algorithm. The trial
wave functions were constructed by solving the equation using the
Thomas--Fermi approximation and multiplying the solution by a phase
factor $\e^{\i l\varphi}$ with a chosen value of $l$. Since states
(\ref{eq:psi}) with different $l$'s are orthogonal, the trial wave
function of this form converges to a state with the same value of
$l$. To solve the GPE system (\ref{eq:full-gpe-dimensionless}),
we used the following trial functions: $\Phi_{3}$ was simply set
to zero, while $\Phi_{1}$ and $\Phi_{2}$ were obtained from the
corresponding dark-state solution $\Phi_{{\rm D}}$ using Eq.~(\ref{eq:ds-comps}).

The chemical potential was calculated from the stationary solution
of the system (\ref{eq:full-gpe-dimensionless}) using
\begin{equation}
\mu=\sum_{i,j}\int\Phi_{i}^{*}(D_{ij}+F_{ij})\Phi_{j}\,\d\boldsymbol{r}\label{eq:mu}
\end{equation}
where
\begin{equation}
\hat{D}=\begin{pmatrix}-\Delta+V & 0 & \Omega_{1}^{*}\\
0 & -\Delta+V & \Omega_{2}^{*}\\
\Omega_{1} & \Omega_{2} & -\Delta+V-\varDelta-\i\frac{\Gamma}{2}
\end{pmatrix}
\end{equation}
and
\begin{equation}
\hat{F}=\begin{pmatrix}g_{11}|\Phi_{1}|^{2}+g_{12}|\Phi_{2}|^{2} & 0 & 0\\
0 & g_{12}|\Phi_{1}|^{2}+g_{22}|\Phi_{2}|^{2} & 0\\
0 & 0 & 0
\end{pmatrix}.
\end{equation}
Meanwhile, the energy per particle $E$ is given by Eq.~(\ref{eq:mu}) with $F_{ij}\to\frac{1}{2}F_{ij}$.

Calculations have been performed using the GPELab software package
\citep{GPELab1,GPELab2}. The figures in the text have been produced using
the Makie.jl software package \citep{Makie.jl}.


\begin{thebibliography}{52}%
\makeatletter
\providecommand \@ifxundefined [1]{%
 \@ifx{#1\undefined}
}%
\providecommand \@ifnum [1]{%
 \ifnum #1\expandafter \@firstoftwo
 \else \expandafter \@secondoftwo
 \fi
}%
\providecommand \@ifx [1]{%
 \ifx #1\expandafter \@firstoftwo
 \else \expandafter \@secondoftwo
 \fi
}%
\providecommand \natexlab [1]{#1}%
\providecommand \enquote  [1]{``#1''}%
\providecommand \bibnamefont  [1]{#1}%
\providecommand \bibfnamefont [1]{#1}%
\providecommand \citenamefont [1]{#1}%
\providecommand \href@noop [0]{\@secondoftwo}%
\providecommand \href [0]{\begingroup \@sanitize@url \@href}%
\providecommand \@href[1]{\@@startlink{#1}\@@href}%
\providecommand \@@href[1]{\endgroup#1\@@endlink}%
\providecommand \@sanitize@url [0]{\catcode `\\12\catcode `\$12\catcode `\&12\catcode `\#12\catcode `\^12\catcode `\_12\catcode `\%12\relax}%
\providecommand \@@startlink[1]{}%
\providecommand \@@endlink[0]{}%
\providecommand \url  [0]{\begingroup\@sanitize@url \@url }%
\providecommand \@url [1]{\endgroup\@href {#1}{\urlprefix }}%
\providecommand \urlprefix  [0]{URL }%
\providecommand \Eprint [0]{\href }%
\providecommand \doibase [0]{https://doi.org/}%
\providecommand \selectlanguage [0]{\@gobble}%
\providecommand \bibinfo  [0]{\@secondoftwo}%
\providecommand \bibfield  [0]{\@secondoftwo}%
\providecommand \translation [1]{[#1]}%
\providecommand \BibitemOpen [0]{}%
\providecommand \bibitemStop [0]{}%
\providecommand \bibitemNoStop [0]{.\EOS\space}%
\providecommand \EOS [0]{\spacefactor3000\relax}%
\providecommand \BibitemShut  [1]{\csname bibitem#1\endcsname}%
\let\auto@bib@innerbib\@empty
\bibitem [{\citenamefont {Fetter}\ and\ \citenamefont {Svidzinsky}(2001)}]{Fetter2001}%
  \BibitemOpen
  \bibfield  {author} {\bibinfo {author} {\bibfnamefont {A.~L.}\ \bibnamefont {Fetter}}\ and\ \bibinfo {author} {\bibfnamefont {A.~A.}\ \bibnamefont {Svidzinsky}},\ }\bibfield  {title} {\bibinfo {title} {Vortices in a trapped dilute {{Bose-Einstein}} condensate},\ }\href {https://doi.org/10.1088/0953-8984/13/12/201} {\bibfield  {journal} {\bibinfo  {journal} {J. Phys.: Condens. Matter}\ }\textbf {\bibinfo {volume} {13}},\ \bibinfo {pages} {R135} (\bibinfo {year} {2001})}\BibitemShut {NoStop}%
\bibitem [{\citenamefont {Tsubota}\ \emph {et~al.}(2013)\citenamefont {Tsubota}, \citenamefont {Kobayashi},\ and\ \citenamefont {Takeuchi}}]{Tsubota2013}%
  \BibitemOpen
  \bibfield  {author} {\bibinfo {author} {\bibfnamefont {M.}~\bibnamefont {Tsubota}}, \bibinfo {author} {\bibfnamefont {M.}~\bibnamefont {Kobayashi}},\ and\ \bibinfo {author} {\bibfnamefont {H.}~\bibnamefont {Takeuchi}},\ }\bibfield  {title} {\bibinfo {title} {Quantum hydrodynamics},\ }\href {https://doi.org/10.1016/j.physrep.2012.09.007} {\bibfield  {journal} {\bibinfo  {journal} {Phys. Rep.}\ }\textbf {\bibinfo {volume} {522}},\ \bibinfo {pages} {191} (\bibinfo {year} {2013})}\BibitemShut {NoStop}%
\bibitem [{\citenamefont {Reeves}\ \emph {et~al.}(2022)\citenamefont {Reeves}, \citenamefont {{Goddard-Lee}}, \citenamefont {Gauthier}, \citenamefont {Stockdale}, \citenamefont {Salman}, \citenamefont {Edmonds}, \citenamefont {Yu}, \citenamefont {Bradley}, \citenamefont {Baker}, \citenamefont {{Rubinsztein-Dunlop}}, \citenamefont {Davis},\ and\ \citenamefont {Neely}}]{Reeves2022}%
  \BibitemOpen
  \bibfield  {author} {\bibinfo {author} {\bibfnamefont {M.~T.}\ \bibnamefont {Reeves}}, \bibinfo {author} {\bibfnamefont {K.}~\bibnamefont {{Goddard-Lee}}}, \bibinfo {author} {\bibfnamefont {G.}~\bibnamefont {Gauthier}}, \bibinfo {author} {\bibfnamefont {O.~R.}\ \bibnamefont {Stockdale}}, \bibinfo {author} {\bibfnamefont {H.}~\bibnamefont {Salman}}, \bibinfo {author} {\bibfnamefont {T.}~\bibnamefont {Edmonds}}, \bibinfo {author} {\bibfnamefont {X.}~\bibnamefont {Yu}}, \bibinfo {author} {\bibfnamefont {A.~S.}\ \bibnamefont {Bradley}}, \bibinfo {author} {\bibfnamefont {M.}~\bibnamefont {Baker}}, \bibinfo {author} {\bibfnamefont {H.}~\bibnamefont {{Rubinsztein-Dunlop}}}, \bibinfo {author} {\bibfnamefont {M.~J.}\ \bibnamefont {Davis}},\ and\ \bibinfo {author} {\bibfnamefont {T.~W.}\ \bibnamefont {Neely}},\ }\bibfield  {title} {\bibinfo {title} {Turbulent relaxation to equilibrium in a two-dimensional quantum vortex gas},\ }\href {https://doi.org/10.1103/PhysRevX.12.011031} {\bibfield  {journal} {\bibinfo
  {journal} {Phys. Rev. X}\ }\textbf {\bibinfo {volume} {12}},\ \bibinfo {pages} {011031} (\bibinfo {year} {2022})}\BibitemShut {NoStop}%
\bibitem [{\citenamefont {Harada}\ \emph {et~al.}(1992)\citenamefont {Harada}, \citenamefont {Matsuda}, \citenamefont {Bonevich}, \citenamefont {Igarashi}, \citenamefont {Kondo}, \citenamefont {Pozzi}, \citenamefont {Kawabe},\ and\ \citenamefont {Tonomura}}]{Harada1992}%
  \BibitemOpen
  \bibfield  {author} {\bibinfo {author} {\bibfnamefont {K.}~\bibnamefont {Harada}}, \bibinfo {author} {\bibfnamefont {T.}~\bibnamefont {Matsuda}}, \bibinfo {author} {\bibfnamefont {J.}~\bibnamefont {Bonevich}}, \bibinfo {author} {\bibfnamefont {M.}~\bibnamefont {Igarashi}}, \bibinfo {author} {\bibfnamefont {S.}~\bibnamefont {Kondo}}, \bibinfo {author} {\bibfnamefont {G.}~\bibnamefont {Pozzi}}, \bibinfo {author} {\bibfnamefont {U.}~\bibnamefont {Kawabe}},\ and\ \bibinfo {author} {\bibfnamefont {A.}~\bibnamefont {Tonomura}},\ }\bibfield  {title} {\bibinfo {title} {Real-time observation of vortex lattices in a superconductor by electron microscopy},\ }\href {https://doi.org/10.1038/360051a0} {\bibfield  {journal} {\bibinfo  {journal} {Nature}\ }\textbf {\bibinfo {volume} {360}},\ \bibinfo {pages} {51} (\bibinfo {year} {1992})}\BibitemShut {NoStop}%
\bibitem [{\citenamefont {Magierski}\ \emph {et~al.}(2024)\citenamefont {Magierski}, \citenamefont {Barresi}, \citenamefont {Makowski}, \citenamefont {P{\k e}cak},\ and\ \citenamefont {Wlaz{\l}owski}}]{Magierski2024}%
  \BibitemOpen
  \bibfield  {author} {\bibinfo {author} {\bibfnamefont {P.}~\bibnamefont {Magierski}}, \bibinfo {author} {\bibfnamefont {A.}~\bibnamefont {Barresi}}, \bibinfo {author} {\bibfnamefont {A.}~\bibnamefont {Makowski}}, \bibinfo {author} {\bibfnamefont {D.}~\bibnamefont {P{\k e}cak}},\ and\ \bibinfo {author} {\bibfnamefont {G.}~\bibnamefont {Wlaz{\l}owski}},\ }\bibfield  {title} {\bibinfo {title} {Quantum vortices in fermionic superfluids: From ultracold atoms to neutron stars.},\ }\href {https://doi.org/10.1140/epja/s10050-024-01378-4} {\bibfield  {journal} {\bibinfo  {journal} {Eur. Phys. J. A}\ }\textbf {\bibinfo {volume} {60}},\ \bibinfo {pages} {186} (\bibinfo {year} {2024})}\BibitemShut {NoStop}%
\bibitem [{\citenamefont {Dobrek}\ \emph {et~al.}(1999)\citenamefont {Dobrek}, \citenamefont {Gajda}, \citenamefont {Lewenstein}, \citenamefont {Sengstock}, \citenamefont {Birkl},\ and\ \citenamefont {Ertmer}}]{Dobrek1999}%
  \BibitemOpen
  \bibfield  {author} {\bibinfo {author} {\bibfnamefont {{\L}.}~\bibnamefont {Dobrek}}, \bibinfo {author} {\bibfnamefont {M.}~\bibnamefont {Gajda}}, \bibinfo {author} {\bibfnamefont {M.}~\bibnamefont {Lewenstein}}, \bibinfo {author} {\bibfnamefont {K.}~\bibnamefont {Sengstock}}, \bibinfo {author} {\bibfnamefont {G.}~\bibnamefont {Birkl}},\ and\ \bibinfo {author} {\bibfnamefont {W.}~\bibnamefont {Ertmer}},\ }\bibfield  {title} {\bibinfo {title} {Optical generation of vortices in trapped {{Bose-Einstein}} condensates},\ }\href {https://doi.org/10.1103/PhysRevA.60.R3381} {\bibfield  {journal} {\bibinfo  {journal} {Phys. Rev. A}\ }\textbf {\bibinfo {volume} {60}},\ \bibinfo {pages} {R3381} (\bibinfo {year} {1999})}\BibitemShut {NoStop}%
\bibitem [{\citenamefont {Madison}\ \emph {et~al.}(2000)\citenamefont {Madison}, \citenamefont {Chevy}, \citenamefont {Wohlleben},\ and\ \citenamefont {Dalibard}}]{Madison2000}%
  \BibitemOpen
  \bibfield  {author} {\bibinfo {author} {\bibfnamefont {K.~W.}\ \bibnamefont {Madison}}, \bibinfo {author} {\bibfnamefont {F.}~\bibnamefont {Chevy}}, \bibinfo {author} {\bibfnamefont {W.}~\bibnamefont {Wohlleben}},\ and\ \bibinfo {author} {\bibfnamefont {J.}~\bibnamefont {Dalibard}},\ }\bibfield  {title} {\bibinfo {title} {Vortex formation in a stirred {{Bose-Einstein}} condensate},\ }\href {https://doi.org/10.1103/PhysRevLett.84.806} {\bibfield  {journal} {\bibinfo  {journal} {Phys. Rev. Lett.}\ }\textbf {\bibinfo {volume} {84}},\ \bibinfo {pages} {806} (\bibinfo {year} {2000})}\BibitemShut {NoStop}%
\bibitem [{\citenamefont {{Abo-Shaeer}}\ \emph {et~al.}(2001)\citenamefont {{Abo-Shaeer}}, \citenamefont {Raman}, \citenamefont {Vogels},\ and\ \citenamefont {Ketterle}}]{AboShaeer2001}%
  \BibitemOpen
  \bibfield  {author} {\bibinfo {author} {\bibfnamefont {J.~R.}\ \bibnamefont {{Abo-Shaeer}}}, \bibinfo {author} {\bibfnamefont {C.}~\bibnamefont {Raman}}, \bibinfo {author} {\bibfnamefont {J.~M.}\ \bibnamefont {Vogels}},\ and\ \bibinfo {author} {\bibfnamefont {W.}~\bibnamefont {Ketterle}},\ }\bibfield  {title} {\bibinfo {title} {Observation of vortex lattices in {{Bose-Einstein}} condensates},\ }\href {https://doi.org/10.1126/science.1060182} {\bibfield  {journal} {\bibinfo  {journal} {Science}\ }\textbf {\bibinfo {volume} {292}},\ \bibinfo {pages} {476} (\bibinfo {year} {2001})}\BibitemShut {NoStop}%
\bibitem [{\citenamefont {Marzlin}\ \emph {et~al.}(1997)\citenamefont {Marzlin}, \citenamefont {Zhang},\ and\ \citenamefont {Wright}}]{Marzlin1997}%
  \BibitemOpen
  \bibfield  {author} {\bibinfo {author} {\bibfnamefont {K.-P.}\ \bibnamefont {Marzlin}}, \bibinfo {author} {\bibfnamefont {W.}~\bibnamefont {Zhang}},\ and\ \bibinfo {author} {\bibfnamefont {E.~M.}\ \bibnamefont {Wright}},\ }\bibfield  {title} {\bibinfo {title} {Vortex coupler for atomic {{Bose-Einstein}} condensates},\ }\href {https://doi.org/10.1103/PhysRevLett.79.4728} {\bibfield  {journal} {\bibinfo  {journal} {Phys. Rev. Lett.}\ }\textbf {\bibinfo {volume} {79}},\ \bibinfo {pages} {4728} (\bibinfo {year} {1997})}\BibitemShut {NoStop}%
\bibitem [{\citenamefont {Dum}\ \emph {et~al.}(1998)\citenamefont {Dum}, \citenamefont {Cirac}, \citenamefont {Lewenstein},\ and\ \citenamefont {Zoller}}]{Dum1998}%
  \BibitemOpen
  \bibfield  {author} {\bibinfo {author} {\bibfnamefont {R.}~\bibnamefont {Dum}}, \bibinfo {author} {\bibfnamefont {J.~I.}\ \bibnamefont {Cirac}}, \bibinfo {author} {\bibfnamefont {M.}~\bibnamefont {Lewenstein}},\ and\ \bibinfo {author} {\bibfnamefont {P.}~\bibnamefont {Zoller}},\ }\bibfield  {title} {\bibinfo {title} {Creation of dark solitons and vortices in {{Bose-Einstein}} condensates},\ }\href {https://doi.org/10.1103/PhysRevLett.80.2972} {\bibfield  {journal} {\bibinfo  {journal} {Phys. Rev. Lett.}\ }\textbf {\bibinfo {volume} {80}},\ \bibinfo {pages} {2972} (\bibinfo {year} {1998})}\BibitemShut {NoStop}%
\bibitem [{\citenamefont {Nandi}\ \emph {et~al.}(2004)\citenamefont {Nandi}, \citenamefont {Walser},\ and\ \citenamefont {Schleich}}]{Nandi2004}%
  \BibitemOpen
  \bibfield  {author} {\bibinfo {author} {\bibfnamefont {G.}~\bibnamefont {Nandi}}, \bibinfo {author} {\bibfnamefont {R.}~\bibnamefont {Walser}},\ and\ \bibinfo {author} {\bibfnamefont {W.~P.}\ \bibnamefont {Schleich}},\ }\bibfield  {title} {\bibinfo {title} {Vortex creation in a trapped {{Bose-Einstein}} condensate by stimulated {{Raman}} adiabatic passage},\ }\href {https://doi.org/10.1103/PhysRevA.69.063606} {\bibfield  {journal} {\bibinfo  {journal} {Phys. Rev. A}\ }\textbf {\bibinfo {volume} {69}},\ \bibinfo {pages} {063606} (\bibinfo {year} {2004})}\BibitemShut {NoStop}%
\bibitem [{\citenamefont {Kapale}\ and\ \citenamefont {Dowling}(2005)}]{Kapale2005}%
  \BibitemOpen
  \bibfield  {author} {\bibinfo {author} {\bibfnamefont {K.~T.}\ \bibnamefont {Kapale}}\ and\ \bibinfo {author} {\bibfnamefont {J.~P.}\ \bibnamefont {Dowling}},\ }\bibfield  {title} {\bibinfo {title} {Vortex phase qubit: {{Generating}} arbitrary, counterrotating, coherent superpositions in {{Bose-Einstein}} condensates via optical angular momentum beams},\ }\href {https://doi.org/10.1103/PhysRevLett.95.173601} {\bibfield  {journal} {\bibinfo  {journal} {Phys. Rev. Lett.}\ }\textbf {\bibinfo {volume} {95}},\ \bibinfo {pages} {173601} (\bibinfo {year} {2005})}\BibitemShut {NoStop}%
\bibitem [{\citenamefont {Juzeli{\=u}nas}\ and\ \citenamefont {{\"O}hberg}(2005)}]{Juzeliunas2005Meissner}%
  \BibitemOpen
  \bibfield  {author} {\bibinfo {author} {\bibfnamefont {G.}~\bibnamefont {Juzeli{\=u}nas}}\ and\ \bibinfo {author} {\bibfnamefont {P.}~\bibnamefont {{\"O}hberg}},\ }\bibfield  {title} {\bibinfo {title} {Creation of an effective magnetic field in ultracold atomic gases using electromagnetically induced transparency},\ }\href {https://doi.org/10.1134/1.2055927} {\bibfield  {journal} {\bibinfo  {journal} {Opt. Spectrosc.}\ }\textbf {\bibinfo {volume} {99}},\ \bibinfo {pages} {357} (\bibinfo {year} {2005})}\BibitemShut {NoStop}%
\bibitem [{\citenamefont {Andersen}\ \emph {et~al.}(2006)\citenamefont {Andersen}, \citenamefont {Ryu}, \citenamefont {Clad{\'e}}, \citenamefont {Natarajan}, \citenamefont {Vaziri}, \citenamefont {Helmerson},\ and\ \citenamefont {Phillips}}]{Andersen2006}%
  \BibitemOpen
  \bibfield  {author} {\bibinfo {author} {\bibfnamefont {M.~F.}\ \bibnamefont {Andersen}}, \bibinfo {author} {\bibfnamefont {C.}~\bibnamefont {Ryu}}, \bibinfo {author} {\bibfnamefont {P.}~\bibnamefont {Clad{\'e}}}, \bibinfo {author} {\bibfnamefont {V.}~\bibnamefont {Natarajan}}, \bibinfo {author} {\bibfnamefont {A.}~\bibnamefont {Vaziri}}, \bibinfo {author} {\bibfnamefont {K.}~\bibnamefont {Helmerson}},\ and\ \bibinfo {author} {\bibfnamefont {W.~D.}\ \bibnamefont {Phillips}},\ }\bibfield  {title} {\bibinfo {title} {Quantized rotation of atoms from photons with orbital angular momentum},\ }\href {https://doi.org/10.1103/PhysRevLett.97.170406} {\bibfield  {journal} {\bibinfo  {journal} {Phys. Rev. Lett.}\ }\textbf {\bibinfo {volume} {97}},\ \bibinfo {pages} {170406} (\bibinfo {year} {2006})}\BibitemShut {NoStop}%
\bibitem [{\citenamefont {Wright}\ \emph {et~al.}(2008)\citenamefont {Wright}, \citenamefont {Leslie},\ and\ \citenamefont {Bigelow}}]{Wright2008}%
  \BibitemOpen
  \bibfield  {author} {\bibinfo {author} {\bibfnamefont {K.~C.}\ \bibnamefont {Wright}}, \bibinfo {author} {\bibfnamefont {L.~S.}\ \bibnamefont {Leslie}},\ and\ \bibinfo {author} {\bibfnamefont {N.~P.}\ \bibnamefont {Bigelow}},\ }\bibfield  {title} {\bibinfo {title} {Optical control of the internal and external angular momentum of a {{Bose-Einstein}} condensate},\ }\href {https://doi.org/10.1103/PhysRevA.77.041601} {\bibfield  {journal} {\bibinfo  {journal} {Phys. Rev. A}\ }\textbf {\bibinfo {volume} {77}},\ \bibinfo {pages} {041601} (\bibinfo {year} {2008})}\BibitemShut {NoStop}%
\bibitem [{\citenamefont {Wright}\ \emph {et~al.}(2009)\citenamefont {Wright}, \citenamefont {Leslie}, \citenamefont {Hansen},\ and\ \citenamefont {Bigelow}}]{Wright2009}%
  \BibitemOpen
  \bibfield  {author} {\bibinfo {author} {\bibfnamefont {K.~C.}\ \bibnamefont {Wright}}, \bibinfo {author} {\bibfnamefont {L.~S.}\ \bibnamefont {Leslie}}, \bibinfo {author} {\bibfnamefont {A.}~\bibnamefont {Hansen}},\ and\ \bibinfo {author} {\bibfnamefont {N.~P.}\ \bibnamefont {Bigelow}},\ }\bibfield  {title} {\bibinfo {title} {Sculpting the vortex state of a spinor {{BEC}}},\ }\href {https://doi.org/10.1103/PhysRevLett.102.030405} {\bibfield  {journal} {\bibinfo  {journal} {Phys. Rev. Lett.}\ }\textbf {\bibinfo {volume} {102}},\ \bibinfo {pages} {030405} (\bibinfo {year} {2009})}\BibitemShut {NoStop}%
\bibitem [{\citenamefont {Mukherjee}\ \emph {et~al.}(2021)\citenamefont {Mukherjee}, \citenamefont {Bandyopadhyay}, \citenamefont {Angom}, \citenamefont {Martin},\ and\ \citenamefont {Majumder}}]{Mukherjee2021}%
  \BibitemOpen
  \bibfield  {author} {\bibinfo {author} {\bibfnamefont {K.}~\bibnamefont {Mukherjee}}, \bibinfo {author} {\bibfnamefont {S.}~\bibnamefont {Bandyopadhyay}}, \bibinfo {author} {\bibfnamefont {D.}~\bibnamefont {Angom}}, \bibinfo {author} {\bibfnamefont {A.~M.}\ \bibnamefont {Martin}},\ and\ \bibinfo {author} {\bibfnamefont {S.}~\bibnamefont {Majumder}},\ }\bibfield  {title} {\bibinfo {title} {Dynamics of the creation of a rotating {{Bose}}--{{Einstein}} condensation by two photon {{Raman}} transition using a {{Laguerre}}--{{Gaussian}} laser pulse},\ }\href {https://doi.org/10.3390/atoms9010014} {\bibfield  {journal} {\bibinfo  {journal} {Atoms}\ }\textbf {\bibinfo {volume} {9}},\ \bibinfo {pages} {14} (\bibinfo {year} {2021})}\BibitemShut {NoStop}%
\bibitem [{\citenamefont {Juzeli{\=u}nas}\ \emph {et~al.}(2007)\citenamefont {Juzeli{\=u}nas}, \citenamefont {Ruseckas}, \citenamefont {{\"O}hberg},\ and\ \citenamefont {Fleischhauer}}]{Juzeliunas2007}%
  \BibitemOpen
  \bibfield  {author} {\bibinfo {author} {\bibfnamefont {G.}~\bibnamefont {Juzeli{\=u}nas}}, \bibinfo {author} {\bibfnamefont {J.}~\bibnamefont {Ruseckas}}, \bibinfo {author} {\bibfnamefont {P.}~\bibnamefont {{\"O}hberg}},\ and\ \bibinfo {author} {\bibfnamefont {M.}~\bibnamefont {Fleischhauer}},\ }\bibfield  {title} {\bibinfo {title} {Formation of solitons in atomic {{Bose}}--{{Einstein}} condensates by dark-state adiabatic passage},\ }\href {https://doi.org/10.3952/lithjphys.47318} {\bibfield  {journal} {\bibinfo  {journal} {Lith. J. Phys.}\ }\textbf {\bibinfo {volume} {47}},\ \bibinfo {pages} {351} (\bibinfo {year} {2007})}\BibitemShut {NoStop}%
\bibitem [{\citenamefont {Gorshkov}\ \emph {et~al.}(2008)\citenamefont {Gorshkov}, \citenamefont {Jiang}, \citenamefont {Greiner}, \citenamefont {Zoller},\ and\ \citenamefont {Lukin}}]{Gorshkov2008}%
  \BibitemOpen
  \bibfield  {author} {\bibinfo {author} {\bibfnamefont {A.~V.}\ \bibnamefont {Gorshkov}}, \bibinfo {author} {\bibfnamefont {L.}~\bibnamefont {Jiang}}, \bibinfo {author} {\bibfnamefont {M.}~\bibnamefont {Greiner}}, \bibinfo {author} {\bibfnamefont {P.}~\bibnamefont {Zoller}},\ and\ \bibinfo {author} {\bibfnamefont {M.~D.}\ \bibnamefont {Lukin}},\ }\bibfield  {title} {\bibinfo {title} {Coherent quantum optical control with subwavelength resolution},\ }\href {https://doi.org/10.1103/PhysRevLett.100.093005} {\bibfield  {journal} {\bibinfo  {journal} {Phys. Rev. Lett.}\ }\textbf {\bibinfo {volume} {100}},\ \bibinfo {pages} {093005} (\bibinfo {year} {2008})}\BibitemShut {NoStop}%
\bibitem [{\citenamefont {{\L}{\k a}cki}\ \emph {et~al.}(2016)\citenamefont {{\L}{\k a}cki}, \citenamefont {Baranov}, \citenamefont {Pichler},\ and\ \citenamefont {Zoller}}]{Lacki2016}%
  \BibitemOpen
  \bibfield  {author} {\bibinfo {author} {\bibfnamefont {M.}~\bibnamefont {{\L}{\k a}cki}}, \bibinfo {author} {\bibfnamefont {M.~A.}\ \bibnamefont {Baranov}}, \bibinfo {author} {\bibfnamefont {H.}~\bibnamefont {Pichler}},\ and\ \bibinfo {author} {\bibfnamefont {P.}~\bibnamefont {Zoller}},\ }\bibfield  {title} {\bibinfo {title} {Nanoscale ``{{Dark State}}'' optical potentials for cold atoms},\ }\href {https://doi.org/10.1103/PhysRevLett.117.233001} {\bibfield  {journal} {\bibinfo  {journal} {Phys. Rev. Lett.}\ }\textbf {\bibinfo {volume} {117}},\ \bibinfo {pages} {233001} (\bibinfo {year} {2016})}\BibitemShut {NoStop}%
\bibitem [{\citenamefont {Gvozdiovas}\ \emph {et~al.}(2023)\citenamefont {Gvozdiovas}, \citenamefont {Spielman},\ and\ \citenamefont {Juzeli{\=u}nas}}]{Gvozdiovas2023}%
  \BibitemOpen
  \bibfield  {author} {\bibinfo {author} {\bibfnamefont {E.}~\bibnamefont {Gvozdiovas}}, \bibinfo {author} {\bibfnamefont {I.~B.}\ \bibnamefont {Spielman}},\ and\ \bibinfo {author} {\bibfnamefont {G.}~\bibnamefont {Juzeli{\=u}nas}},\ }\bibfield  {title} {\bibinfo {title} {Interference-induced anisotropy in a two-dimensional dark-state optical lattice},\ }\href {https://doi.org/10.1103/PhysRevA.107.033328} {\bibfield  {journal} {\bibinfo  {journal} {Phys. Rev. A}\ }\textbf {\bibinfo {volume} {107}},\ \bibinfo {pages} {033328} (\bibinfo {year} {2023})}\BibitemShut {NoStop}%
\bibitem [{\citenamefont {Hamedi}\ \emph {et~al.}(2022)\citenamefont {Hamedi}, \citenamefont {{\v Z}labys}, \citenamefont {Ahufinger}, \citenamefont {Halfmann}, \citenamefont {Mompart},\ and\ \citenamefont {Juzeli{\=u}nas}}]{Hamedi2022}%
  \BibitemOpen
  \bibfield  {author} {\bibinfo {author} {\bibfnamefont {H.~R.}\ \bibnamefont {Hamedi}}, \bibinfo {author} {\bibfnamefont {G.}~\bibnamefont {{\v Z}labys}}, \bibinfo {author} {\bibfnamefont {V.}~\bibnamefont {Ahufinger}}, \bibinfo {author} {\bibfnamefont {T.}~\bibnamefont {Halfmann}}, \bibinfo {author} {\bibfnamefont {J.}~\bibnamefont {Mompart}},\ and\ \bibinfo {author} {\bibfnamefont {G.}~\bibnamefont {Juzeli{\=u}nas}},\ }\bibfield  {title} {\bibinfo {title} {Spatially strongly confined atomic excitation via a two dimensional stimulated {{Raman}} adiabatic passage},\ }\href {https://doi.org/10.1364/OE.447397} {\bibfield  {journal} {\bibinfo  {journal} {Opt. Express}\ }\textbf {\bibinfo {volume} {30}},\ \bibinfo {pages} {13915} (\bibinfo {year} {2022})}\BibitemShut {NoStop}%
\bibitem [{\citenamefont {Dutton}\ and\ \citenamefont {Ruostekoski}(2004)}]{Dutton2004}%
  \BibitemOpen
  \bibfield  {author} {\bibinfo {author} {\bibfnamefont {Z.}~\bibnamefont {Dutton}}\ and\ \bibinfo {author} {\bibfnamefont {J.}~\bibnamefont {Ruostekoski}},\ }\bibfield  {title} {\bibinfo {title} {Transfer and storage of vortex states in light and matter waves},\ }\href {https://doi.org/10.1103/PhysRevLett.93.193602} {\bibfield  {journal} {\bibinfo  {journal} {Phys. Rev. Lett.}\ }\textbf {\bibinfo {volume} {93}},\ \bibinfo {pages} {193602} (\bibinfo {year} {2004})}\BibitemShut {NoStop}%
\bibitem [{\citenamefont {Thanvanthri}\ \emph {et~al.}(2008)\citenamefont {Thanvanthri}, \citenamefont {Kapale},\ and\ \citenamefont {Dowling}}]{Thanvanthri2008}%
  \BibitemOpen
  \bibfield  {author} {\bibinfo {author} {\bibfnamefont {S.}~\bibnamefont {Thanvanthri}}, \bibinfo {author} {\bibfnamefont {K.~T.}\ \bibnamefont {Kapale}},\ and\ \bibinfo {author} {\bibfnamefont {J.~P.}\ \bibnamefont {Dowling}},\ }\bibfield  {title} {\bibinfo {title} {Arbitrary coherent superpositions of quantized vortices in {{Bose-Einstein}} condensates via orbital angular momentum of light},\ }\href {https://doi.org/10.1103/PhysRevA.77.053825} {\bibfield  {journal} {\bibinfo  {journal} {Phys. Rev. A}\ }\textbf {\bibinfo {volume} {77}},\ \bibinfo {pages} {053825} (\bibinfo {year} {2008})}\BibitemShut {NoStop}%
\bibitem [{\citenamefont {Dum}\ and\ \citenamefont {Olshanii}(1996)}]{Dum1996}%
  \BibitemOpen
  \bibfield  {author} {\bibinfo {author} {\bibfnamefont {R.}~\bibnamefont {Dum}}\ and\ \bibinfo {author} {\bibfnamefont {M.}~\bibnamefont {Olshanii}},\ }\bibfield  {title} {\bibinfo {title} {Gauge structures in atom-laser interaction: {{Bloch}} oscillations in a dark lattice},\ }\href {https://doi.org/10.1103/PhysRevLett.76.1788} {\bibfield  {journal} {\bibinfo  {journal} {Phys. Rev. Lett.}\ }\textbf {\bibinfo {volume} {76}},\ \bibinfo {pages} {1788} (\bibinfo {year} {1996})}\BibitemShut {NoStop}%
\bibitem [{\citenamefont {Juzeli{\=u}nas}\ and\ \citenamefont {{\"O}hberg}(2004)}]{Juzeliunas2004}%
  \BibitemOpen
  \bibfield  {author} {\bibinfo {author} {\bibfnamefont {G.}~\bibnamefont {Juzeli{\=u}nas}}\ and\ \bibinfo {author} {\bibfnamefont {P.}~\bibnamefont {{\"O}hberg}},\ }\bibfield  {title} {\bibinfo {title} {Slow light in degenerate {{Fermi}} gases},\ }\href {https://doi.org/10.1103/PhysRevLett.93.033602} {\bibfield  {journal} {\bibinfo  {journal} {Phys. Rev. Lett.}\ }\textbf {\bibinfo {volume} {93}},\ \bibinfo {pages} {033602} (\bibinfo {year} {2004})}\BibitemShut {NoStop}%
\bibitem [{\citenamefont {Juzeli{\=u}nas}\ \emph {et~al.}(2005{\natexlab{a}})\citenamefont {Juzeli{\=u}nas}, \citenamefont {{\"O}hberg}, \citenamefont {Ruseckas},\ and\ \citenamefont {Klein}}]{Juzeliunas2005}%
  \BibitemOpen
  \bibfield  {author} {\bibinfo {author} {\bibfnamefont {G.}~\bibnamefont {Juzeli{\=u}nas}}, \bibinfo {author} {\bibfnamefont {P.}~\bibnamefont {{\"O}hberg}}, \bibinfo {author} {\bibfnamefont {J.}~\bibnamefont {Ruseckas}},\ and\ \bibinfo {author} {\bibfnamefont {A.}~\bibnamefont {Klein}},\ }\bibfield  {title} {\bibinfo {title} {Effective magnetic fields in degenerate atomic gases induced by light beams with orbital angular momenta},\ }\href {https://doi.org/10.1103/PhysRevA.71.053614} {\bibfield  {journal} {\bibinfo  {journal} {Phys. Rev. A}\ }\textbf {\bibinfo {volume} {71}},\ \bibinfo {pages} {053614} (\bibinfo {year} {2005}{\natexlab{a}})}\BibitemShut {NoStop}%
\bibitem [{\citenamefont {Juzeli{\=u}nas}\ \emph {et~al.}(2005{\natexlab{b}})\citenamefont {Juzeli{\=u}nas}, \citenamefont {Ruseckas},\ and\ \citenamefont {{\"O}hberg}}]{Juzeliunas2005EIT}%
  \BibitemOpen
  \bibfield  {author} {\bibinfo {author} {\bibfnamefont {G.}~\bibnamefont {Juzeli{\=u}nas}}, \bibinfo {author} {\bibfnamefont {J.}~\bibnamefont {Ruseckas}},\ and\ \bibinfo {author} {\bibfnamefont {P.}~\bibnamefont {{\"O}hberg}},\ }\bibfield  {title} {\bibinfo {title} {Effective magnetic fields induced by {{EIT}} in ultra-cold atomic gases},\ }\href {https://doi.org/10.1088/0953-4075/38/23/001} {\bibfield  {journal} {\bibinfo  {journal} {J. Phys. B: At. Mol. Opt. Phys.}\ }\textbf {\bibinfo {volume} {38}},\ \bibinfo {pages} {4171} (\bibinfo {year} {2005}{\natexlab{b}})}\BibitemShut {NoStop}%
\bibitem [{\citenamefont {Jendrzejewski}\ \emph {et~al.}(2016)\citenamefont {Jendrzejewski}, \citenamefont {Eckel}, \citenamefont {Tiecke}, \citenamefont {Juzeli{\=u}nas}, \citenamefont {Campbell}, \citenamefont {Jiang},\ and\ \citenamefont {Gorshkov}}]{Jendrzejewski2016}%
  \BibitemOpen
  \bibfield  {author} {\bibinfo {author} {\bibfnamefont {F.}~\bibnamefont {Jendrzejewski}}, \bibinfo {author} {\bibfnamefont {S.}~\bibnamefont {Eckel}}, \bibinfo {author} {\bibfnamefont {T.~G.}\ \bibnamefont {Tiecke}}, \bibinfo {author} {\bibfnamefont {G.}~\bibnamefont {Juzeli{\=u}nas}}, \bibinfo {author} {\bibfnamefont {G.~K.}\ \bibnamefont {Campbell}}, \bibinfo {author} {\bibfnamefont {L.}~\bibnamefont {Jiang}},\ and\ \bibinfo {author} {\bibfnamefont {A.~V.}\ \bibnamefont {Gorshkov}},\ }\bibfield  {title} {\bibinfo {title} {Subwavelength-width optical tunnel junctions for ultracold atoms},\ }\href {https://doi.org/10.1103/PhysRevA.94.063422} {\bibfield  {journal} {\bibinfo  {journal} {Phys. Rev. A}\ }\textbf {\bibinfo {volume} {94}},\ \bibinfo {pages} {063422} (\bibinfo {year} {2016})}\BibitemShut {NoStop}%
\bibitem [{\citenamefont {Scully}\ and\ \citenamefont {Zubairy}(1997)}]{ScullyZubairyBook}%
  \BibitemOpen
  \bibfield  {author} {\bibinfo {author} {\bibfnamefont {M.~O.}\ \bibnamefont {Scully}}\ and\ \bibinfo {author} {\bibfnamefont {M.~S.}\ \bibnamefont {Zubairy}},\ }\href {https://doi.org/10.1017/CBO9780511813993} {\emph {\bibinfo {title} {Quantum {{Optics}}}}}\ (\bibinfo  {publisher} {Cambridge University Press},\ \bibinfo {address} {Cambridge},\ \bibinfo {year} {1997})\BibitemShut {NoStop}%
\bibitem [{\citenamefont {Pitaevskii}\ and\ \citenamefont {Stringari}(2016)}]{PitaevskiiStringari}%
  \BibitemOpen
  \bibfield  {author} {\bibinfo {author} {\bibfnamefont {L.~P.}\ \bibnamefont {Pitaevskii}}\ and\ \bibinfo {author} {\bibfnamefont {S.}~\bibnamefont {Stringari}},\ }\href {https://doi.org/10.1093/acprof:oso/9780198758884.001.0001} {\emph {\bibinfo {title} {Bose-{{Einstein}} Condensation and Superfluidity}}}\ (\bibinfo  {publisher} {Oxford University Press},\ \bibinfo {address} {Oxford},\ \bibinfo {year} {2016})\BibitemShut {NoStop}%
\bibitem [{\citenamefont {Gaunt}\ \emph {et~al.}(2013)\citenamefont {Gaunt}, \citenamefont {Schmidutz}, \citenamefont {Gotlibovych}, \citenamefont {Smith},\ and\ \citenamefont {Hadzibabic}}]{Gaunt2013}%
  \BibitemOpen
  \bibfield  {author} {\bibinfo {author} {\bibfnamefont {A.~L.}\ \bibnamefont {Gaunt}}, \bibinfo {author} {\bibfnamefont {T.~F.}\ \bibnamefont {Schmidutz}}, \bibinfo {author} {\bibfnamefont {I.}~\bibnamefont {Gotlibovych}}, \bibinfo {author} {\bibfnamefont {R.~P.}\ \bibnamefont {Smith}},\ and\ \bibinfo {author} {\bibfnamefont {Z.}~\bibnamefont {Hadzibabic}},\ }\bibfield  {title} {\bibinfo {title} {Bose-{{Einstein}} condensation of atoms in a uniform potential},\ }\href {https://doi.org/10.1103/PhysRevLett.110.200406} {\bibfield  {journal} {\bibinfo  {journal} {Phys. Rev. Lett.}\ }\textbf {\bibinfo {volume} {110}},\ \bibinfo {pages} {200406} (\bibinfo {year} {2013})}\BibitemShut {NoStop}%
\bibitem [{\citenamefont {Gutterres}\ \emph {et~al.}(2002)\citenamefont {Gutterres}, \citenamefont {Amiot}, \citenamefont {Fioretti}, \citenamefont {Gabbanini}, \citenamefont {Mazzoni},\ and\ \citenamefont {Dulieu}}]{Gutterres2013}%
  \BibitemOpen
  \bibfield  {author} {\bibinfo {author} {\bibfnamefont {R.~F.}\ \bibnamefont {Gutterres}}, \bibinfo {author} {\bibfnamefont {C.}~\bibnamefont {Amiot}}, \bibinfo {author} {\bibfnamefont {A.}~\bibnamefont {Fioretti}}, \bibinfo {author} {\bibfnamefont {C.}~\bibnamefont {Gabbanini}}, \bibinfo {author} {\bibfnamefont {M.}~\bibnamefont {Mazzoni}},\ and\ \bibinfo {author} {\bibfnamefont {O.}~\bibnamefont {Dulieu}},\ }\bibfield  {title} {\bibinfo {title} {Determination of the ${}^{87}\mathrm{Rb}$ $5p$ state dipole matrix element and radiative lifetime from the photoassociation spectroscopy of the $\mathrm{Rb}_{2}$ ${0}_{g}^{\ensuremath{-}}{(P}_{3/2})$ long-range state},\ }\href {https://doi.org/10.1103/PhysRevA.66.024502} {\bibfield  {journal} {\bibinfo  {journal} {Phys. Rev. A}\ }\textbf {\bibinfo {volume} {66}},\ \bibinfo {pages} {024502} (\bibinfo {year} {2002})}\BibitemShut {NoStop}%
\bibitem [{\citenamefont {Egorov}\ \emph {et~al.}(2013)\citenamefont {Egorov}, \citenamefont {Opanchuk}, \citenamefont {Drummond}, \citenamefont {Hall}, \citenamefont {Hannaford},\ and\ \citenamefont {Sidorov}}]{Egorov2013}%
  \BibitemOpen
  \bibfield  {author} {\bibinfo {author} {\bibfnamefont {M.}~\bibnamefont {Egorov}}, \bibinfo {author} {\bibfnamefont {B.}~\bibnamefont {Opanchuk}}, \bibinfo {author} {\bibfnamefont {P.}~\bibnamefont {Drummond}}, \bibinfo {author} {\bibfnamefont {B.~V.}\ \bibnamefont {Hall}}, \bibinfo {author} {\bibfnamefont {P.}~\bibnamefont {Hannaford}},\ and\ \bibinfo {author} {\bibfnamefont {A.~I.}\ \bibnamefont {Sidorov}},\ }\bibfield  {title} {\bibinfo {title} {Measurement of $s$-wave scattering lengths in a two-component {Bose-Einstein} condensate},\ }\href {https://doi.org/10.1103/PhysRevA.87.053614} {\bibfield  {journal} {\bibinfo  {journal} {Phys. Rev. A}\ }\textbf {\bibinfo {volume} {87}},\ \bibinfo {pages} {053614} (\bibinfo {year} {2013})}\BibitemShut {NoStop}%
\bibitem [{\citenamefont {Patrick}\ \emph {et~al.}(2023)\citenamefont {Patrick}, \citenamefont {Gupta}, \citenamefont {Gregory},\ and\ \citenamefont {Barenghi}}]{Patrick2023}%
  \BibitemOpen
  \bibfield  {author} {\bibinfo {author} {\bibfnamefont {S.}~\bibnamefont {Patrick}}, \bibinfo {author} {\bibfnamefont {A.}~\bibnamefont {Gupta}}, \bibinfo {author} {\bibfnamefont {R.}~\bibnamefont {Gregory}},\ and\ \bibinfo {author} {\bibfnamefont {C.~F.}\ \bibnamefont {Barenghi}},\ }\bibfield  {title} {\bibinfo {title} {Stability of quantized vortices in two-component condensates},\ }\href {https://doi.org/10.1103/PhysRevResearch.5.033201} {\bibfield  {journal} {\bibinfo  {journal} {Phys. Rev. Res.}\ }\textbf {\bibinfo {volume} {5}},\ \bibinfo {pages} {033201} (\bibinfo {year} {2023})}\BibitemShut {NoStop}%
\bibitem [{\citenamefont {Richaud}\ \emph {et~al.}(2023)\citenamefont {Richaud}, \citenamefont {Lamporesi}, \citenamefont {Capone},\ and\ \citenamefont {Recati}}]{Richaud2023}%
  \BibitemOpen
  \bibfield  {author} {\bibinfo {author} {\bibfnamefont {A.}~\bibnamefont {Richaud}}, \bibinfo {author} {\bibfnamefont {G.}~\bibnamefont {Lamporesi}}, \bibinfo {author} {\bibfnamefont {M.}~\bibnamefont {Capone}},\ and\ \bibinfo {author} {\bibfnamefont {A.}~\bibnamefont {Recati}},\ }\bibfield  {title} {\bibinfo {title} {Mass-driven vortex collisions in flat superfluids},\ }\href {https://doi.org/10.1103/PhysRevA.107.053317} {\bibfield  {journal} {\bibinfo  {journal} {Phys. Rev. A}\ }\textbf {\bibinfo {volume} {107}},\ \bibinfo {pages} {053317} (\bibinfo {year} {2023})}\BibitemShut {NoStop}%
\bibitem [{\citenamefont {Ao}\ and\ \citenamefont {Chui}(1998)}]{Ao1998}%
  \BibitemOpen
  \bibfield  {author} {\bibinfo {author} {\bibfnamefont {P.}~\bibnamefont {Ao}}\ and\ \bibinfo {author} {\bibfnamefont {S.~T.}\ \bibnamefont {Chui}},\ }\bibfield  {title} {\bibinfo {title} {Binary {{Bose-Einstein}} condensate mixtures in weakly and strongly segregated phases},\ }\href {https://doi.org/10.1103/PhysRevA.58.4836} {\bibfield  {journal} {\bibinfo  {journal} {Phys. Rev. A}\ }\textbf {\bibinfo {volume} {58}},\ \bibinfo {pages} {4836} (\bibinfo {year} {1998})}\BibitemShut {NoStop}%
\bibitem [{\citenamefont {Timmermans}(1998)}]{Timmermans1998}%
  \BibitemOpen
  \bibfield  {author} {\bibinfo {author} {\bibfnamefont {E.}~\bibnamefont {Timmermans}},\ }\bibfield  {title} {\bibinfo {title} {Phase separation of {{Bose-Einstein}} condensates},\ }\href {https://doi.org/10.1103/PhysRevLett.81.5718} {\bibfield  {journal} {\bibinfo  {journal} {Phys. Rev. Lett.}\ }\textbf {\bibinfo {volume} {81}},\ \bibinfo {pages} {5718} (\bibinfo {year} {1998})}\BibitemShut {NoStop}%
\bibitem [{\citenamefont {Wen}\ \emph {et~al.}(2012)\citenamefont {Wen}, \citenamefont {Liu}, \citenamefont {Cai}, \citenamefont {Zhang},\ and\ \citenamefont {Hu}}]{Wen2012}%
  \BibitemOpen
  \bibfield  {author} {\bibinfo {author} {\bibfnamefont {L.}~\bibnamefont {Wen}}, \bibinfo {author} {\bibfnamefont {W.~M.}\ \bibnamefont {Liu}}, \bibinfo {author} {\bibfnamefont {Y.}~\bibnamefont {Cai}}, \bibinfo {author} {\bibfnamefont {J.~M.}\ \bibnamefont {Zhang}},\ and\ \bibinfo {author} {\bibfnamefont {J.}~\bibnamefont {Hu}},\ }\bibfield  {title} {\bibinfo {title} {Controlling phase separation of a two-component {{Bose-Einstein}} condensate by confinement},\ }\href {https://doi.org/10.1103/PhysRevA.85.043602} {\bibfield  {journal} {\bibinfo  {journal} {Phys. Rev. A}\ }\textbf {\bibinfo {volume} {85}},\ \bibinfo {pages} {043602} (\bibinfo {year} {2012})}\BibitemShut {NoStop}%
\bibitem [{\citenamefont {Di}\ \emph {et~al.}(2023)\citenamefont {Di}, \citenamefont {Nie},\ and\ \citenamefont {Yang}}]{Di2023}%
  \BibitemOpen
  \bibfield  {author} {\bibinfo {author} {\bibfnamefont {X.}~\bibnamefont {Di}}, \bibinfo {author} {\bibfnamefont {Y.-H.}\ \bibnamefont {Nie}},\ and\ \bibinfo {author} {\bibfnamefont {T.}~\bibnamefont {Yang}},\ }\bibfield  {title} {\bibinfo {title} {Manipulating vortices with a rotating laser beam in {{Bose}}--{{Einstein}} condensates},\ }\href {https://doi.org/10.1088/1555-6611/acde6d} {\bibfield  {journal} {\bibinfo  {journal} {Laser Phys.}\ }\textbf {\bibinfo {volume} {33}},\ \bibinfo {pages} {085501} (\bibinfo {year} {2023})}\BibitemShut {NoStop}%
\bibitem [{\citenamefont {Pethick}\ and\ \citenamefont {Smith}(2008)}]{PethickSmith}%
  \BibitemOpen
  \bibfield  {author} {\bibinfo {author} {\bibfnamefont {C.~J.}\ \bibnamefont {Pethick}}\ and\ \bibinfo {author} {\bibfnamefont {H.}~\bibnamefont {Smith}},\ }\href {https://doi.org/10.1017/CBO9780511802850} {\emph {\bibinfo {title} {Bose--{{Einstein}} Condensation in Dilute Gases}}},\ \bibinfo {edition} {2nd}\ ed.\ (\bibinfo  {publisher} {Cambridge University Press},\ \bibinfo {address} {Cambridge},\ \bibinfo {year} {2008})\BibitemShut {NoStop}%
\bibitem [{\citenamefont {Raman}\ \emph {et~al.}(1999)\citenamefont {Raman}, \citenamefont {K{\"o}hl}, \citenamefont {Onofrio}, \citenamefont {Durfee}, \citenamefont {Kuklewicz}, \citenamefont {Hadzibabic},\ and\ \citenamefont {Ketterle}}]{Raman1999}%
  \BibitemOpen
  \bibfield  {author} {\bibinfo {author} {\bibfnamefont {C.}~\bibnamefont {Raman}}, \bibinfo {author} {\bibfnamefont {M.}~\bibnamefont {K{\"o}hl}}, \bibinfo {author} {\bibfnamefont {R.}~\bibnamefont {Onofrio}}, \bibinfo {author} {\bibfnamefont {D.~S.}\ \bibnamefont {Durfee}}, \bibinfo {author} {\bibfnamefont {C.~E.}\ \bibnamefont {Kuklewicz}}, \bibinfo {author} {\bibfnamefont {Z.}~\bibnamefont {Hadzibabic}},\ and\ \bibinfo {author} {\bibfnamefont {W.}~\bibnamefont {Ketterle}},\ }\bibfield  {title} {\bibinfo {title} {Evidence for a critical velocity in a {{Bose-Einstein}} condensed gas},\ }\href {https://doi.org/10.1103/PhysRevLett.83.2502} {\bibfield  {journal} {\bibinfo  {journal} {Phys. Rev. Lett.}\ }\textbf {\bibinfo {volume} {83}},\ \bibinfo {pages} {2502} (\bibinfo {year} {1999})}\BibitemShut {NoStop}%
\bibitem [{\citenamefont {Onofrio}\ \emph {et~al.}(2000)\citenamefont {Onofrio}, \citenamefont {Raman}, \citenamefont {Vogels}, \citenamefont {{Abo-Shaeer}}, \citenamefont {Chikkatur},\ and\ \citenamefont {Ketterle}}]{Onofrio2000}%
  \BibitemOpen
  \bibfield  {author} {\bibinfo {author} {\bibfnamefont {R.}~\bibnamefont {Onofrio}}, \bibinfo {author} {\bibfnamefont {C.}~\bibnamefont {Raman}}, \bibinfo {author} {\bibfnamefont {J.~M.}\ \bibnamefont {Vogels}}, \bibinfo {author} {\bibfnamefont {J.~R.}\ \bibnamefont {{Abo-Shaeer}}}, \bibinfo {author} {\bibfnamefont {A.~P.}\ \bibnamefont {Chikkatur}},\ and\ \bibinfo {author} {\bibfnamefont {W.}~\bibnamefont {Ketterle}},\ }\bibfield  {title} {\bibinfo {title} {Observation of superfluid flow in a {{Bose-Einstein}} condensed gas},\ }\href {https://doi.org/10.1103/PhysRevLett.85.2228} {\bibfield  {journal} {\bibinfo  {journal} {Phys. Rev. Lett.}\ }\textbf {\bibinfo {volume} {85}},\ \bibinfo {pages} {2228} (\bibinfo {year} {2000})}\BibitemShut {NoStop}%
\bibitem [{\citenamefont {Desbuquois}\ \emph {et~al.}(2012)\citenamefont {Desbuquois}, \citenamefont {Chomaz}, \citenamefont {Yefsah}, \citenamefont {L{\'e}onard}, \citenamefont {Beugnon}, \citenamefont {Weitenberg},\ and\ \citenamefont {Dalibard}}]{Desbuquois2012}%
  \BibitemOpen
  \bibfield  {author} {\bibinfo {author} {\bibfnamefont {R.}~\bibnamefont {Desbuquois}}, \bibinfo {author} {\bibfnamefont {L.}~\bibnamefont {Chomaz}}, \bibinfo {author} {\bibfnamefont {T.}~\bibnamefont {Yefsah}}, \bibinfo {author} {\bibfnamefont {J.}~\bibnamefont {L{\'e}onard}}, \bibinfo {author} {\bibfnamefont {J.}~\bibnamefont {Beugnon}}, \bibinfo {author} {\bibfnamefont {C.}~\bibnamefont {Weitenberg}},\ and\ \bibinfo {author} {\bibfnamefont {J.}~\bibnamefont {Dalibard}},\ }\bibfield  {title} {\bibinfo {title} {Superfluid behaviour of a two-dimensional {{Bose}} gas},\ }\href {https://doi.org/10.1038/nphys2378} {\bibfield  {journal} {\bibinfo  {journal} {Nat. Phys.}\ }\textbf {\bibinfo {volume} {8}},\ \bibinfo {pages} {645} (\bibinfo {year} {2012})}\BibitemShut {NoStop}%
\bibitem [{\citenamefont {Weimer}\ \emph {et~al.}(2015)\citenamefont {Weimer}, \citenamefont {Morgener}, \citenamefont {Singh}, \citenamefont {Siegl}, \citenamefont {Hueck}, \citenamefont {Luick}, \citenamefont {Mathey},\ and\ \citenamefont {Moritz}}]{Weimer2015}%
  \BibitemOpen
  \bibfield  {author} {\bibinfo {author} {\bibfnamefont {W.}~\bibnamefont {Weimer}}, \bibinfo {author} {\bibfnamefont {K.}~\bibnamefont {Morgener}}, \bibinfo {author} {\bibfnamefont {V.~P.}\ \bibnamefont {Singh}}, \bibinfo {author} {\bibfnamefont {J.}~\bibnamefont {Siegl}}, \bibinfo {author} {\bibfnamefont {K.}~\bibnamefont {Hueck}}, \bibinfo {author} {\bibfnamefont {N.}~\bibnamefont {Luick}}, \bibinfo {author} {\bibfnamefont {L.}~\bibnamefont {Mathey}},\ and\ \bibinfo {author} {\bibfnamefont {H.}~\bibnamefont {Moritz}},\ }\bibfield  {title} {\bibinfo {title} {Critical velocity in the {{BEC-BCS Crossover}}},\ }\href {https://doi.org/10.1103/PhysRevLett.114.095301} {\bibfield  {journal} {\bibinfo  {journal} {Phys. Rev. Lett.}\ }\textbf {\bibinfo {volume} {114}},\ \bibinfo {pages} {095301} (\bibinfo {year} {2015})}\BibitemShut {NoStop}%
\bibitem [{\citenamefont {Singh}\ \emph {et~al.}(2016)\citenamefont {Singh}, \citenamefont {Weimer}, \citenamefont {Morgener}, \citenamefont {Siegl}, \citenamefont {Hueck}, \citenamefont {Luick}, \citenamefont {Moritz},\ and\ \citenamefont {Mathey}}]{Singh2016}%
  \BibitemOpen
  \bibfield  {author} {\bibinfo {author} {\bibfnamefont {V.~P.}\ \bibnamefont {Singh}}, \bibinfo {author} {\bibfnamefont {W.}~\bibnamefont {Weimer}}, \bibinfo {author} {\bibfnamefont {K.}~\bibnamefont {Morgener}}, \bibinfo {author} {\bibfnamefont {J.}~\bibnamefont {Siegl}}, \bibinfo {author} {\bibfnamefont {K.}~\bibnamefont {Hueck}}, \bibinfo {author} {\bibfnamefont {N.}~\bibnamefont {Luick}}, \bibinfo {author} {\bibfnamefont {H.}~\bibnamefont {Moritz}},\ and\ \bibinfo {author} {\bibfnamefont {L.}~\bibnamefont {Mathey}},\ }\bibfield  {title} {\bibinfo {title} {Probing superfluidity of {{Bose-Einstein}} condensates via laser stirring},\ }\href {https://doi.org/10.1103/PhysRevA.93.023634} {\bibfield  {journal} {\bibinfo  {journal} {Phys. Rev. A}\ }\textbf {\bibinfo {volume} {93}},\ \bibinfo {pages} {023634} (\bibinfo {year} {2016})}\BibitemShut {NoStop}%
\bibitem [{\citenamefont {Singh}\ \emph {et~al.}(2017)\citenamefont {Singh}, \citenamefont {Weitenberg}, \citenamefont {Dalibard},\ and\ \citenamefont {Mathey}}]{Singh2017}%
  \BibitemOpen
  \bibfield  {author} {\bibinfo {author} {\bibfnamefont {V.~P.}\ \bibnamefont {Singh}}, \bibinfo {author} {\bibfnamefont {C.}~\bibnamefont {Weitenberg}}, \bibinfo {author} {\bibfnamefont {J.}~\bibnamefont {Dalibard}},\ and\ \bibinfo {author} {\bibfnamefont {L.}~\bibnamefont {Mathey}},\ }\bibfield  {title} {\bibinfo {title} {Superfluidity and relaxation dynamics of a laser-stirred two-dimensional {{Bose}} gas},\ }\href {https://doi.org/10.1103/PhysRevA.95.043631} {\bibfield  {journal} {\bibinfo  {journal} {Phys. Rev. A}\ }\textbf {\bibinfo {volume} {95}},\ \bibinfo {pages} {043631} (\bibinfo {year} {2017})}\BibitemShut {NoStop}%
\bibitem [{\citenamefont {Kiehn}\ \emph {et~al.}(2022)\citenamefont {Kiehn}, \citenamefont {Singh},\ and\ \citenamefont {Mathey}}]{Kiehn2022}%
  \BibitemOpen
  \bibfield  {author} {\bibinfo {author} {\bibfnamefont {H.}~\bibnamefont {Kiehn}}, \bibinfo {author} {\bibfnamefont {V.~P.}\ \bibnamefont {Singh}},\ and\ \bibinfo {author} {\bibfnamefont {L.}~\bibnamefont {Mathey}},\ }\bibfield  {title} {\bibinfo {title} {Superfluidity of a laser-stirred {{Bose-Einstein}} condensate},\ }\href {https://doi.org/10.1103/PhysRevA.105.043317} {\bibfield  {journal} {\bibinfo  {journal} {Phys. Rev. A}\ }\textbf {\bibinfo {volume} {105}},\ \bibinfo {pages} {043317} (\bibinfo {year} {2022})}\BibitemShut {NoStop}%
\bibitem [{\citenamefont {Astrakharchik}\ and\ \citenamefont {Pitaevskii}(2004)}]{Astrakharchik2004}%
  \BibitemOpen
  \bibfield  {author} {\bibinfo {author} {\bibfnamefont {G.~E.}\ \bibnamefont {Astrakharchik}}\ and\ \bibinfo {author} {\bibfnamefont {L.~P.}\ \bibnamefont {Pitaevskii}},\ }\bibfield  {title} {\bibinfo {title} {Motion of a heavy impurity through a {{Bose-Einstein}} condensate},\ }\href {https://doi.org/10.1103/PhysRevA.70.013608} {\bibfield  {journal} {\bibinfo  {journal} {Phys. Rev. A}\ }\textbf {\bibinfo {volume} {70}},\ \bibinfo {pages} {013608} (\bibinfo {year} {2004})}\BibitemShut {NoStop}%
\bibitem [{\citenamefont {Antoine}\ and\ \citenamefont {Duboscq}(2014)}]{GPELab1}%
  \BibitemOpen
  \bibfield  {author} {\bibinfo {author} {\bibfnamefont {X.}~\bibnamefont {Antoine}}\ and\ \bibinfo {author} {\bibfnamefont {R.}~\bibnamefont {Duboscq}},\ }\bibfield  {title} {\bibinfo {title} {{{GPELab}}, a {{Matlab}} toolbox to solve {{Gross}}--{{Pitaevskii}} equations {{I}}: {{Computation}} of stationary solutions},\ }\href {https://doi.org/10.1016/j.cpc.2014.06.026} {\bibfield  {journal} {\bibinfo  {journal} {Comput. Phys. Commun.}\ }\textbf {\bibinfo {volume} {185}},\ \bibinfo {pages} {2969} (\bibinfo {year} {2014})}\BibitemShut {NoStop}%
\bibitem [{\citenamefont {Antoine}\ and\ \citenamefont {Duboscq}(2015)}]{GPELab2}%
  \BibitemOpen
  \bibfield  {author} {\bibinfo {author} {\bibfnamefont {X.}~\bibnamefont {Antoine}}\ and\ \bibinfo {author} {\bibfnamefont {R.}~\bibnamefont {Duboscq}},\ }\bibfield  {title} {\bibinfo {title} {{{GPELab}}, a {{Matlab}} toolbox to solve {{Gross}}--{{Pitaevskii}} equations {{II}}: {{Dynamics}} and stochastic simulations},\ }\href {https://doi.org/10.1016/j.cpc.2015.03.012} {\bibfield  {journal} {\bibinfo  {journal} {Comput. Phys. Commun.}\ }\textbf {\bibinfo {volume} {193}},\ \bibinfo {pages} {95} (\bibinfo {year} {2015})}\BibitemShut {NoStop}%
\bibitem [{\citenamefont {Danisch}\ and\ \citenamefont {Krumbiegel}(2021)}]{Makie.jl}%
  \BibitemOpen
  \bibfield  {author} {\bibinfo {author} {\bibfnamefont {S.}~\bibnamefont {Danisch}}\ and\ \bibinfo {author} {\bibfnamefont {J.}~\bibnamefont {Krumbiegel}},\ }\bibfield  {title} {\bibinfo {title} {Makie.jl: {{Flexible}} high-performance data visualization for {{Julia}}},\ }\href {https://doi.org/10.21105/joss.03349} {\bibfield  {journal} {\bibinfo  {journal} {J. Open Source Soft.}\ }\textbf {\bibinfo {volume} {6}},\ \bibinfo {pages} {3349} (\bibinfo {year} {2021})}\BibitemShut {NoStop}%
\end{thebibliography}
\end{document}